\documentclass[conference]{IEEEtran}
\usepackage{ifpdf}
\usepackage{longtable}
\usepackage{multirow}

\usepackage{color}

\usepackage{cite}
\ifCLASSINFOpdf
   \usepackage[pdftex]{graphicx}
\else
   \usepackage[dvips]{graphicx}
   \graphicspath{{../eps/}}
\fi
\usepackage[cmex10]{amsmath}
\usepackage{hyperref}
\usepackage{array}
\usepackage{mdwmath}
\usepackage{mdwtab}
\usepackage[font=footnotesize]{subfig}
\begin{document}
\title{Analytical Models for Energy Consumption in Infrastructure WLAN STAs Carrying TCP Traffic}

% \author{\IEEEauthorblockN{Pranav Agrawal}
% \IEEEauthorblockA{CEDT\\
% Indian Institute of Science\\
% Bangalore, India 560012\\
% Email: pagarwal@cedt.iisc.ernet.in}
% \and
% \IEEEauthorblockN{Anurag Kumar}
% \IEEEauthorblockA{ECE\\
% Indian Institute of Science\\
% Bangalore, India 560012\\
% Email: anurag@ece.iisc.ernet.in}
% \and
% \IEEEauthorblockN{Joy Kuri}
% \IEEEauthorblockA{CEDT\\
% Indian Institute of Science\\
% Bangalore, India 560012\\
% Email: kuri@cedt.iisc.ernet.in}}


\author{\IEEEauthorblockN{Pranav Agrawal\IEEEauthorrefmark{1},
A. Kumar\IEEEauthorrefmark{2},
J. Kuri\IEEEauthorrefmark{1}, 
M. Panda\IEEEauthorrefmark{2},
V. Navda\IEEEauthorrefmark{3},
R. Ramjee\IEEEauthorrefmark{3},
V. N. Padmanabhan\IEEEauthorrefmark{3}
\IEEEauthorblockA{\IEEEauthorrefmark{1}Centre for Electronics Design and Technology }
\IEEEauthorblockA{\IEEEauthorrefmark{2}Electrical and Communication Engineering}
Indian Institute of Science, Bangalore, India - 560 012
\IEEEauthorblockA{\IEEEauthorrefmark{3}Microsoft Research, Bangalore, India}}}
\maketitle
\begin{abstract}
%\boldmath
We develop analytical models for estimating the energy spent by stations (STAs) in infrastructure WLANs when performing TCP controlled file downloads. We focus on the energy spent in radio communication when the STAs are in the Continuously Active Mode (CAM), or in the static Power Save Mode (PSM). Our approach is to develop accurate models for obtaining the fraction of times the STA radios spend in idling, receiving and transmitting. We discuss two traffic models for each mode of operation: (i) each STA performs one large file download, and (ii) the STAs perform short file transfers.  We evaluate the rate of STA energy expenditure with long file downloads, and show that static PSM is worse than just using CAM. For short file downloads we compute the number of file downloads that can be completed with given battery capacity, and show that PSM performs better than CAM for this case. We provide a validation of our analytical models using the NS-2 simulator~\cite{simulator:ns_2}.
 \end{abstract}
% \fi
\IEEEpeerreviewmaketitle
\section{Introduction}
With the advent of wireless technology in hand-held devices, saving energy incurred due to wireless protocols is of prime importance. To design efficient power management policies for wireless adapters, it is often required to know the energy spent by wireless stations (STAs) running different classes of applications. Since users often use TCP based applications, in this paper we characterize the energy spent by STAs while running TCP controlled data transfers. We focus only on the energy spent in radio communication, and we evaluate this by obtaining the fraction of times the STA radio stays in different states, i.e., idle, transmit and receive.

In normal mode of operation, also called as the Continuously Active Mode (CAM), an STA always keeps its radio on, so it can receive and transmit at any time. This mode of operation is energy inefficient since STAs draw current even when they are idling. To save power during the period when there is less or no network activity, WiFi cards are provided with controls through which they can be turned off. To leverage this facility, the IEEE 802.11 standard has a feature using which STAs can turn off their radio without losing packets. This is generally the Power Save Mode (PSM). In this mode, an STA can be in any one of the two state, \emph{active state} and \emph{sleep state}.

\emph{Contribution:}
In this paper we analyze various scenarios in which several stations (STAs) are associated with a single Access Point (AP). In each scenario each STA is considered to be either in the PSM or the CAM and downloading files via the AP. The file server is considered to be located on the high speed Ethernet link connected to the AP, which makes the propagation delay between the AP and server negligible. Our aim is to model the energy spent for radio communication by the STAs for the following two types of TCP traffic in either mode of operation of the STA:

\begin{itemize}
 \item  $N$ STAs downloading long files over TCP -- In this scenario, the average rate of expenditure of energy is analyzed.
To evaluate this, we obtain the fraction of times the radios of STAs stay in different states, i.e., idle, receive and transmit. 

\item $N$ STAs downloading short files over TCP -- In this scenario, we consider a constant number of users downloading short files over TCP. In between two downloads, a short period of inactivity or \emph{think time}. To analyze this scenario, a Processor Sharing (PS) model is used to model the download rates provided to the files by IEEE 802.11 MAC. The PS service rate is obtained from the analysis of long file downloads mentioned in the previous paragraph.
\end{itemize}

This paper is organized as follows: Section \ref{sec:related_work} discusses the previous literature in the area of WLAN energy modeling. Section \ref{sec:psm_overview} provides an overview of the PSM and the queuing structure at the AP.  In Section \ref{sec:long_file}, we analyze the scenario of TCP long transfers for both CAM and PSM. In Section~\ref{sec:short_files}, we analyze short file transfers. Finally, Section~\ref{sec:conclusion} concludes the paper.
 
\section{Related Work}\label{sec:related_work}
Anastasi et al.~\cite{Analytical:saving_energy_psm_pspoll_anastasi_04} consider a single STA in PSM downloading a file over TCP in the presence of $N$ active STAs. The authors evaluate the expected energy spent by the STA to download the file as a function of  $N$. This is indirectly obtained by evaluating the contention time required to send a PS-POLL frame. While in our work, we do not consider active STAs, instead they are considered to be downloading files over TCP. Further, we have assumed more realistic PSM protocol, which will be explained later.

Lei and Nilsson~\cite{Analytical:bulk_service_lei_nilsson_05} consider STAs in PSM carrying downlink traffic in which inter-arrival time between packets at the AP is exponentially distributed. They obtained average Packet Delay due to queueing and the PSM protocol. They also obtain lower and upper bounds on the average percentage of time a STA stays in sleep state.  Baek and Choi \cite{Analytical:perform_extension_of_nelson_09} consider the same scenario as in~\cite{Analytical:bulk_service_lei_nilsson_05} and evaluate the variance and exact expression for the average percentage of time a STA stays in sleep state. 
Si et al.~\cite{Analytical:novel_down_link_scheme_pengbo_08} consider STAs in PSM mode, carrying downlink and uplink traffic. They describes the model, using which aggregate throughput and power consumption is obtained. We note that the models in~\cite{Analytical:bulk_service_lei_nilsson_05},~\cite{Analytical:perform_extension_of_nelson_09} and~\cite{Analytical:novel_down_link_scheme_pengbo_08} assume Poisson arrivals of packets into the AP or STA, and, hence, do not correctly model the traffic generated by the TCP controlled file downloads.
 They observed excessive contention among STAs immediately after the transmission of beacon frame, if there are large number of STAs in PSM. It is because, STAs start contending for PS-POLL frame safter receiving beacon frame, if they have a data frame stored packet at the AP. To avoid excessive contention, authors suggested that the AP should not inform to all STAs about the stored packets. This is also observed by us and can drastically reduce the performance of PSM. However, in our case, we avoid excessive contention by setting the rule at the STA; if the STA has already sent the PS-POLL and is waiting for the packet, it will not generate another PS-POLL before receiving the packet, even if beacon frame indicates that there are packets stored at the AP.

In all the above papers \cite{Analytical:saving_energy_psm_pspoll_anastasi_04},~\cite{Analytical:bulk_service_lei_nilsson_05},~\cite{Analytical:perform_extension_of_nelson_09} and~\cite{Analytical:novel_down_link_scheme_pengbo_08}, authors consider PSM protocol implementation which is not practical in the presence of download type background traffic. They consider the following sequence of frame exchanges: First the PSM STA sends the PS-POLL frame through contention, after SIFS AP sends the data packet and after SIFS again the STA sends the MAC ACK. So the AP does not contend to send data. In the presence of traffic from the AP to other STAs, when the AP receives the PS-POLL frame, some packets might be already present in the NIC queue of the AP, and these packets need to be sent first. We have analyzed the PSM protocol in the presence of traffic from other STAs, which is a more realistic scenario and so we have considered an implementation of the PSM protocol in which the AP contends to send data to PSM STAs.
 Hu et al.~\cite{Analytical:micro_ibss_rong_06} consider STAs in an independent basic service set (IBSS) and evaluate the throughput, delay and the loss rate of the energy characterization as a function of the traffic load, buffer size and other protocol specific parameters. Our work is different from this as we focus on the STAs in an Infrastructure Basic Service Set.

Krashinsky and Balakrishnan~\cite {Experimental:ronny_bsd_05}  consider a single STA in PSM doing very short file transfers (order of tens of KBs). They observed that web transfers incur large delays, because of the interaction between TCP slow start, RTT and PSM. To bound the delay, the authors propose a bounded slow down (BSD) protocol in which a web page can experience a delay not more than a specified percentage ($p$) of the actual normal delay (without PSM). BSD~\cite {Experimental:ronny_bsd_05} is further improved upon by Quiao and Shin~\cite{Experimental:smart_psm_quaio_05}. They estimates the RTT of current TCP connection and using this information, the sleep wake schedule is made more efficient. However, the scope of their work is limited to only one STA, while here, we analyze the effect of background traffic, which plays a dominant role in determining energy consumption and delay for a file transfer. 

 Anand et al.~\cite{Experimental:selftuning_night_03} demonstrate the degradation of performance of latency sensitive application like NFS and audio streaming for a STA in PSM. To prevent this degradation, they proposed a self-tuning power management (STPM) algorithm  which accepts inputs from applications.  On the basis of the inputs, the algorithm evaluates expected energy and delay incurred in both the modes (CAM and PSM), using which it decides to operate in a particular mode. Our work can complement this by quantifying the exact value of the energy that is consumed while a TCP application is running, so it can help to devise better power management policies. 
Yong et al.~\cite{Experimental:scheduled_psm_yong_07} propose a way to minimize energy and delay by scheduling and informing the  schedule to STAs through beacon frames. 
 The authors show that scheduled PSM improves the performance in terms of energy as it reduces the idle times in the presence of background traffic.
Tan et al.~\cite{Experimental:psm_throttling_enhua_07} propose to take advantage of throttling done by the TCP server in media streaming applications. 
 Throttling means that the server sends data at rate less than the end to end available bandwidth. 
 They use TCP receiver advertised window to shape traffic in the form of periodic bursts. So instead of downloading packets over larger duration, an STA completes the download in lesser duration, which saves energy. However if the WLAN is the bottleneck link then it is difficult to achieve to achieve this gain.
Zanella and Pellegrini et al.~\cite{Analytical:math_analysis_dcf_04} and Wang et al.~\cite{Analytical:CAM_Nnodes_dharma_06} consider $N$ saturated STAs in CAM mode and analyze the energy spent by them in radio communication. While in our paper, STAs are not considered to be saturated, but they are considered to be downloading files over TCP.
Baiamonte et al.~\cite{Experimental:saving_energy_biamonte_06} propose to make use of NAV set in RTS and CTS, using which the non intended receiver can switch to low power state during the upcoming transmission. 
 It requires that switching delay to low power state should be less than the transmission time of the data packet.  Our model also accounts for the time during which an STA listens to the traffic for other STAs.
\section{PSM - Overview} \label{sec:psm_overview}

There are some situations which are not specified in the protocol but are implementation dependent. Such situations and the assumed behaviors of an STA and the AP are described here.
After sending a PS-POLL, the STA marks its state as \emph{waiting for unicast}. If before the STA receives the unicast packet, the AP transmits a beacon frame and it indicates that there are packets at the AP for this STA, then this STA will not generate another PS-POLL frame. But this may result in a deadlock when the packet that it is waiting for is lost, because then the STA will continue to be awake and will not send another PS-POLL. To prevent this situation, a timer is started when the STA sends the PS-POLL, and if the STA does not receive a packet before timer expiry, it goes to the sleep state. Subsequently in the next beacon interval, if the STA gets an indication, then it will send a PS-POLL to retrieve the packet from the AP. Further, if the beacon frames arrives at the STA when it is contending for PS-POLL, then it ignores the beacon frame, because the STA already knows that there is a packet at the AP for it.
% % \setlength{\abovecaptionskip}{0pt}
% \setlength{\floatsep}{0pt}
 \setlength{\textfloatsep}{0pt}
\begin{figure}
\centering
\includegraphics[width= 2.5 in]{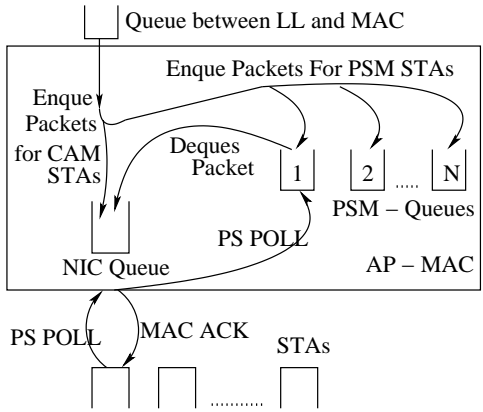} 
\caption{Queuing Structure at AP}
\label{fig:psm_queue}
\end{figure}
\section{Modeling Assumptions}
In this section we state the assumptions common to all the scenarios we have analyzed. We consider a single cell 802.11 WLAN with $N$ STAs associated with a single AP. The STAs and the AP contend for the channel via the DCF mechanism. We consider various scenarios in which either all the STAs are in CAM or all in PSM.  Here, we consider only TCP controlled download traffic, which means that the AP sends data packets to STAs, while STAs send TCP ACKs. We assume that the RTS/CTS mechanism is used by the AP to send data packets, while the basic access scheme is used by the STAs to send TCP ACKs. The following are our modeling assumptions:
\begin{itemize}
\item In all the scenarios, at any instant, an STA has at most a single TCP connection.
\item The application at the STA is such that flow control is never required and the advertised TCP window is always $W_{\max}$.
\item The receivers do not implement the \emph{delayed TCP ACK} strategy, i.e., every received packet generates a TCP ACK.
\item The file server is assumed to be connected to the AP by a high speed LAN, which implies that the propagation delay between the AP and the TCP server can be neglected.
\item The buffers are large enough so that there is no loss of packets due to buffer overflow at the AP or STAs.  
\item There are no packet losses due to the bit errors on the the wireless medium. Also, there are no packet drops due to the excessive collisions in the medium. The analysis can be extended to include packet loss due to bit errors; further for TCP controlled transfers the collision probability is indeed so low that the packets are rarely dropped at the MAC layer due to excessive collisions.
\item We also assume that when there are $k$ active STAs, then these STAs and the AP attempt in any slot with probability $\beta_{k+1}$, where $\beta_{k+1}$ is long term attempt rate and is obtained via saturation analysis in \cite{Base:kumar_fpa_05}.
\end{itemize}
\section{Long File Transfer} \label{sec:long_file}
In this section, we consider $N$ STAs associated with the AP, with each one downloading a single large file over TCP. We consider two scenarios: 1) $N$ STAs in CAM, 2) $N$ STAs in PSM. For both the scenarios, we obtain expressions for throughput and average current drawn as a function of the number of STAs.
\subsection{All STAs in CAM} \label{sec:long_files_CAM}
 Let $X_{ack}(t)$ be the total number of ACKs stored in all the STAs at any instant $t$, $X_{data}(t)$ be the number of data packets stored at AP at $t$. Since the RTT between the AP and the server is negligible, so a data packet arrives immediately after the arrival of the TCP ACK at the AP. By assumption, $W_{\max}$ is the TCP window advertised by the receiver, so at any instant, $X_{ack}(t) + X_{data}(t) = NW_{\max}$. Which implies it is sufficient to keep track of either $X_{ack}(t)$ or $X_{data}(t)$. In the model, we assume that TCP ACKs are uniformly distributed among STAs, which is quite a valid assumption as there is no preference given to any STA. The model here we use is the simplified version of the model described in \cite{Base:bcg_tcp_throughput_06}, in which the authors consider both upload and download traffic, to evaluate the aggregate throughput. In the next section, we develop a new model for calculating energy expenditure rates.

\begin{figure}[b]
\begin{center}
 \scalebox{.4}{ \input{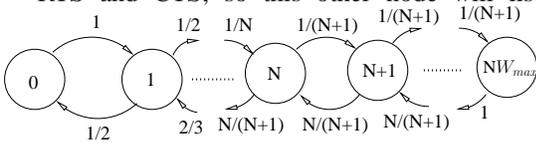}}  
 \caption{DTMC of the process $X_k$}
\label{fig:markovchain_exact}
\end{center}
\end{figure}

Let us call the instants just after successful transmission of a packet on the medium, as the \emph{success instants}, and denote the $k^{th}$ success instant as $G_k$. Let the value of $X_{ack}(t)$ at instant $G_k$ be $X_k$. Since, we are approximating IEEE 802.11 MAC by $p$-persistent model, in which every wireless entity attempts independently in every slot with probability $\beta_k$, where $k$ is the number of active entities. Because of it, given the state of $X(t)$ at $G_k$, the future evolution of the process is independent of the past. Under the above assumptions, $\{(X_k;G_k),k \geq 0\}$ forms a Markov renewal sequence, and process $X(t)$ forms a Markov regenerative process. The DTMC of the process $X_k$ is shown in the Figure~\ref{fig:markovchain_exact}. A transition from state $i$ to $i+1$ represents the success of the AP and a transition from state $i$ to $i-1$ represents success of some STA. Since the backoff parameters of all the STAs and the AP are same, if $X_k =i$, then at the next \emph{success instant} the AP wins the contention with probability $1/(\min(N,i) + 1)$ and one of the $\min(i,N)$ STA wins with the probability $\min(i,N)/(\min(i,N)+1)$.
% \begin{figure}
%   \centering
%  \includegraphics[width
% = 3.5 in]{./eps/embedded_markovchain_exact.eps}
% \caption{Embedded Markov Chain}\label{fig:markovchain_exact}
% \end{figure}
\subsubsection{Aggregate Throughput}
Consider the process $X_k$ and define $T_k = G_{k+1} - G_{k}$ as the length of the $k^{th}$ cycle. Let the number of successful attempts by the AP in the $k^{th}$ cycle be denoted by $H_k$; $H_k$ can be either $1$ or $0$. Let $H(t)$ denote the number of successful attempts made by the AP in $(0,t)$. Then by Markov regenerative analysis~\cite{book:book_rw_wolff}, the following can be written:
\begin{align}
\begin{split}
 \Theta_N &= \lim_{t \to \infty }\frac{H(t)}{t} = \frac{\sum_{k = 0}^{N} \frac{1}{k+1} \pi_k + \sum_{k = N+1}^{NW_{max}-1} \frac{1}{N+1} \pi_k }{\sum_{k = 0}^{NW_{max}}\pi_{k} E_k[T]}
\end{split}
\end{align}
where, $\pi_k$ is the stationary probability of $k$ STAs contending in a cycle and $E_k[X]$ is the expected time until the end of the next success, when the number of STA at the beginning is $k$.
The detailed expression for $E_k[T]$ is given in Appendix~\ref{appendix:mean_cycle_length_long_cam}.

\subsubsection{Average Current}
In this section, expressions for the average current drawn by an STA is evaluated. For this, we obtain the fraction of time any STA spends in different communication states. We define the following possible states:
\begin{itemize}
\item Transmitting State (Tx): In this state, the STA is transmitting. 
\item Receiving State (Rx): In this state, the STA is receiving.  However, there could be two substates corresponding to this state.
\begin{itemize}
\item Receiving Decode state (RxD): In this state, the STA is receiving as well as decoding.
\item Receiving Listen state (RxLs): In this state, the STA is  receiving but not decoding the data. This state is possible because of channel reservation by RTS-CTS mechanism. If the channel is reserved for two nodes, than any other node will know the length of the reservation from the "Duration" field in the RTS and CTS; so this other node will listen to the ongoing transfer and can choose not to decode the corresponding packets. This can result in less consumption of power than in the receive decode state~\cite{tech_spec:atheros_card_2state_power}.
\end{itemize}
\item Idle State (Id) - In this state, the channel is idle; no node is transmitting.
\item Sleep State (Sl) - In this state, STA is in sleep state and draws a very small current.
\end{itemize}
Let us denote the above states as $M_1 = Tx$, $M_2 = RxD$, $M_3 = RxLs$, $M_4 = Id$, $M_5 = Sl$. 
Let us denote $J_{M_r}$ as the current drawn by an STA while it is in state $M_r$. Let us define $Q_k(t)$ as the total charge drawn by STA $k$ in the time interval ($0,t$), then the average current ($J_{av}$) drawn by STAs can be written as follows:
\begin{align}
 J_{av} &= \frac{1}{N}\sum_{k=1}^{N}\lim_{t \to \infty}\frac{Q_k(t)}{t} 
\label{eqn:avg_curr}
\end{align}
Let us define the following indicator functions for an STA $k$:
\begin{align}
  I_{M_r}^k(u) = \left\{ \begin{array}{cc}
         1 & \mbox{ if  STA $k$ is in state $M_r$ at instant $u$}\\
         0 & \mbox{otherwise}.\end{array} \right. 
\label{eqn:indicator_fun}
\end{align}
Now, writing $Q_k(t)$ in terms of the above indicator functions:
\begin{align}
Q_k(t) = \sum_{r=1}^{5}J_{M_r}\int_{0}^{t} I_{M_r}^k(u)\,\,du 
\label{eqn:long_file_q_k_t}
\end{align}
% \begin{align}
% \begin{split}
%  J_{av} &= \lim_{t \to \infty}\frac{1}{t}\sum_{r=1}^{5}J_{M_r}\int_{0}^{t} I_{M_r}^k(u)\,\,du \\
% 	&= \sum_{r=1}^{5} J_{M_r}\lim_{t \to \infty}\frac{1}{t}\int_{0}^{t} I_{M_r}^k(u)\,\,du \\
% &=J_{M_r}\Phi_{M_r}
% \label{eqn:avg_curr}
% \end{split}
% \end{align}
By substituting, Eqn.~\ref{eqn:long_file_q_k_t} in Eqn.~\ref{eqn:avg_curr},and then rearranging it, we get the following equation for average current:
\begin{align}
\begin{split}
 J_{av} &= \frac{1}{N}\sum_{k=1}^{N}\lim_{t \to \infty}\frac{1}{t}\sum_{r=1}^{5}J_{M_r}\int_{0}^{t} I_{M_r}^k(u)\,\,du \\
&= \sum_{r=1}^{5} J_{M_r}\frac{1}{N}\lim_{t \to \infty}\sum_{k=1}^{N}\frac{1}{t}\int_{0}^{t} I_{M_r}^k(u)\,\,du  \\
&= \sum_{r=1}^{5} J_{M_r}\Phi_{M_r}
\label{eqn:avg_curr_1}
\end{split}
\end{align}
where, $\Phi_{M_r}$ as the fraction of time an STA spends in state $M_r$. The finite sum and the limit can be exchanged so the above rearrangement is valid. Our aim is to evaluate $\Phi_{M_r}$: 
\begin{align}
\begin{split}
\Phi_{M_r} &=\frac{1}{N}\lim_{t \to \infty}\sum_{k=1}^{N}\frac{1}{t}\int_{0}^{t} I_{M_r}^k(u)\,\,du \\
&=\frac{1}{N}\lim_{t \to \infty}\frac{1}{t}\int_0^t   S_{M_r}(u) du 
\label{eqn:fraction_time}
\end{split}
\end{align}
where,
\begin{equation}
S_{M_r}(u)= \sum_{k=1}^{N} I_{M_r}^k(u)
\label{eqn:long_file_S} 
\end{equation}
Then by Markov regenerative analysis~\cite{book:book_rw_wolff}, one can show the following,
\begin{align}
\begin{split} 
&\lim_{t \to \infty} \frac{1}{t}\int_{0}^t S_{M_r}(u) du = \frac{\sum_{k=0}^{NW_{max}} \pi_k E_{k}\left[ \int_{G_{k-1}}^{G_{k}} S_{M_r}(u)\,du\right]}{\sum_{k=0}^{NW_{max}} \pi_k E_k[T]} \\
\end{split}
\end{align}
The detailed expression for $E_k[T]$ is given in Appendix~\ref{appendix:mean_cycle_length_long_cam}, and that for $E_{k}\left[ \int_{G_{k-1}}^{G_{k}} S_{M_r}(u)\,du\right]$ in Appendix~\ref{appendix:fracs_times_CAM_long}.
\subsection {All STAs in PSM}
\subsubsection{$N>5$ - Development of the Model} \label{sec:long_files_PSM_N5}
In this section, we consider $N$ STAs in PSM, downloading large files over TCP. In this scenario, AP will always contend for the channel, since for every two packets ($1$ TCP ACK + $1$ PS-POLL) sent by each of the STAs, $N$ packets need to be transmitted by the AP. Since, no preference is given either to the AP or to the STA, the above situation is possible, only if small number of STAs contend at any time, so that $2/3$ of the packets are transmitted by the STAs and $1/3$ of them are transmitted by the AP. Since, we are assuming negligible RTT between the AP and server, so at any time, most of the packets of the TCP windows of the STAs are present at the AP. 
Because of this the "More" bit is always set in every data packet sent; so the STAs never go to sleep. 
On receiving a packet, the STA has to send a PS-POLL frame and a TCP ACK. Since the PS-POLL is a MAC level packet, it is enqueued at the HOL position in the transmission queue of the STA and the TCP ACK at the end of the queue. If the transmission queue of the STA is empty when it receives the data packet, then immediately after the reception, its transmission queue will contain two packets, PS-POLL at the HOL position and TCP-ACK behind the PS-POLL. After STA sends a PS-POLL it starts contending for TCP-ACK.
If the STA queue is nonempty (it implies that the STA is contending for TCP ACK) when it receives the data packet, then the STA will first transmit the PS-POLL. To transmit the PS-POLL, the STA will not sample the new backoff, but uses the residual backoff of the TCP ACK for which it was already contending when it received the data packet. It is not possible that STA receives the data packet when it is contending for PS-POLL, because it is only after the PS-POLL is sent a data packet arrives at the STA. 
% Recalling, the beacon frames are ignored, if the STA is waiting for the packet or contending for PS-POLL, so no extra PS-POLL are generated. 

When the AP receives a PS-POLL packet from the STA, then a data packet corresponding to this STA is brought into the NIC or transmission queue of the AP. There might be some packets already in the transmission queue of the AP (the probability of this increases with $N$), due to which this packet will be transmitted only after the packets preceding it are transmitted. During the time when the AP transmits these preceding packets, with high probability, the STA will transmit the TCP ACK; as a result the AP always sends a data packet to an STA that has an empty transmission queue.

Since no preference is given to the AP and the AP sends a single packet per PS-POLL, the transmission queue of the AP will build up for large value of $N$.
The following can be inferred on the basis of the above discussion:
1) A packet successfully transmitted by the AP goes to an empty STA and the total number of contending STAs increases by one;
2) There are some STAs that are contending to send PS-POLLs and some are contending to send TCP-ACKs;
3) When a STA successfully transmits a PS-POLL, the number of STAs contending for TCP-ACK increases by one;
4) When a STA successfully transmits a TCP ACK, the number of STAs contending decreases by one.

\begin{figure}[htbp]
\begin{center}
 \scalebox{.3}{ \input{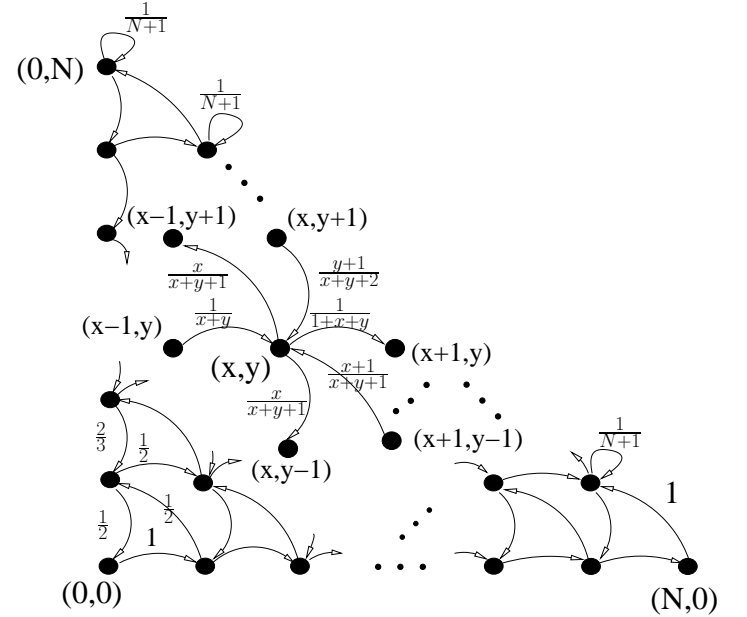} }  
 \caption{2-Dimensional DTMC of the process $(X_k,Y_k)$ }
\label{fig:2_D_markov}
\end{center}
\end{figure}

Consider the process $X(t)$ of the number of STAs with a PS-POLL at the HOL position and TCP ACK behind it, and the process $Y(t)$ of the number of STAs with only a TCP ACK. 
Consider the joint process $(X(t),Y(t))$, embed it at the ends of success instants. Let us denote $G_k$ the instant when the $k^{th}$ successful transmission ends. Let us denote $(X_k,Y_k)$ as the value of the process $(X(t),Y(t))$ at $G_k$. Define $T_k = G_{k+1} - G_{k}$. Using the same arguments of $p$-persistent approximation, as stated earlier in the Section~\ref{sec:long_files_CAM}, $\{(X_k,Y_k);G_k , k\geq 0\}$ forms a Markov renewal sequence, and the process $(X(t),Y(t))$ forms a Markov regenerative process. The transition probabilities of the Markov chain of $(X_k,Y_k)$ depend on the number of active STAs, and are shown in Fig. \ref{fig:2_D_markov}. In Fig.~\ref{fig:2_D_markov}, the $x$ axis represents the process $X_k$ and the $y$ axis represents process $Y_k$, and the state space is given by $\{(x,y):0\leq x+y \leq N\}$. The transition probabilities are obtained using the fact that all nodes (Wireless entities in WLAN) have equal chance to transmit; so the transition probability from $(x_1,y_1)$ to $(x_1+1,y_1)$, which corresponds a successful transmission by the AP, is given by $1/(x_1 + y_1+1)$. Other transition probabilities are also obtained in the same way.
\subsubsection{$N > 5$ - Aggregate download throughput}
Let the number of successful attempts by the AP in the $k^{th}$ cycle be denoted by $H_k$; it could be either 1 or 0. Let $H(t)$ denote the number of successful attempts made by the AP in $(0,t)$. By Markov regenerative analysis~\cite{book:book_rw_wolff} the following can be written,
\begin{align}
\begin{split}
 \Theta_N &= \lim_{t \to \infty }\frac{H(t)}{t} \\&= \frac{\sum_{j = 0}^{N - 1}\sum_{i = 0}^{N - j - 1} \pi_{i,j} \frac{1}{i + j + 1}+ \sum_{i,j:i+j=N,i\neq N}\pi_{i,j}\frac{1}{N+1}}{\sum_{j = 0}^{N}\sum_{i = 0}^{N}\pi_{i,j} E_{i,j}[T]}
\end{split}
\end{align}
% \begin{align}
% \begin{split}
%  \Theta_N &= \lim_{t \to \infty }\frac{H(t)}{t} \\&= \frac{\sum_{j = 0}^{N - 1}\sum_{i = 0}^{N - j - 1} \pi_{i,j}p_{i,j}^{i+1,j} + \sum_{i,j:i+j=N,i\neq N}\pi_{i,j}p_{i,j}^{i,j}}{\sum_{j = 0}^{N}\sum_{i = 0}^{N}\pi_{i,j} E_{i,j}[T]}
% \end{split}
% \end{align}
$\pi_{i,j}$ is the stationary probability of the process $(X_k,Y_k)$. \\
$E_{i,j}[T]$ is the expected time until the next success, starting with the state $(i,j)$ and its detailed derivation is given in Appendix~\ref{appendix:mean_cycle_length_long_psm_N>5}.
% \begin{figure}
%   \centering
%  \includegraphics[width
% = 2.5 in]{./eps/plots_long_tcp_psm/throughput_psm/thr_02_55_11.eps}
% \caption{Aggregate Throughput - TCP Download (PSM)}
%   \label{fig:throughput_long_tcp_psm}
% \end{figure}
\subsubsection{$N = 1$ - Aggregate download throughput}
 After the Slow Start phase is over and the TCP window has grown to its maximum value, there will always be some packets at the AP and STA, with high probability. Due to this, the AP will always set the More bit in every outgoing packet; so STA will never go to sleep.

When an STA receives a packet with the More bit set, it has to send a PS-POLL and a TCP ACK. PS-POLL, being MAC level packet, will be enqueued at the HOL position of the NIC queue, while the TCP ACK is enqueued at the end of the queue. 
Since after the transmission of packet the AP queue is empty, so transmission of PS-POLL occurs without contention. However, TCP ACKs and data packets contend for transmission.

Consider the process $X(t)$ denoting the number of TCP ACKs with the STA. The number of data packets with the AP is  $W - X(t)$. Denote the end of the $k^{th}$ success instants as $G_k$. Let $X_k$ be the number of TCP ACKs with the STA at $G_k$. Let $T_k = G_{k+1} - G_k$. Let the number of successful attempts by AP be $H(t)$ in time interval $(0,t)$. The number of successful attempts by the AP is 0 or 1 in between $G_k$ and $G_{k+1}$ with probability $0.5$. Then, using the Renewal Reward Theorem, the  following can be written,
\begin{equation}
 \Theta_1 = \lim_{t \to \infty} \frac{H(t)}{t}= \frac{0.5}{E[T_k]}
\end{equation}
where, the detailed expression for $E[T]$ is given in Appendix~\ref{appendix:mean_cycle_length_long_psm_N1}. 
\subsubsection{Average Current with N STAs in PSM}
Equations~\ref{eqn:avg_curr}--~\ref{eqn:long_file_S} remain valid for this scenario also. The expression for various fractions is given by the following equation:
\begin{align}
\begin{split} 
&\Phi_{M_r} = \frac{1}{N}\lim_{t \to \infty} \frac{1}{t}\int_{0}^t S_{M_r}(u) du \\&= \frac{1}{N}\frac{\sum_{i=0}^{N} \sum_{j=0}^{N} \pi_{i,j} E_{i,j}\left[\int_{G_{k-1}}^{G_{k}} S_{M_r}(u)du\right]}{\sum_{i=0}^{N}\sum_{j=0}^{N} \pi_{i,j} E_{i,j}[T]} 
\end{split}
\end{align}
The detailed expression for $E_{i,j}[T]$ is given in Appendix~\ref{appendix:mean_cycle_length_long_psm_N>5} and that for $E_{i,j}\left[\int_{G_{k-1}}^{G_{k}} S_{W_r}(u)du\right]$ in Appendix~\ref{appendix:fracs_times_PSM_long}. Model for one STA PSM downloading large file ($N=1$) is shown in Appendix~\ref{appendix:avg_current_psm_N1}.
\begin{figure*}
  \centering
  \subfloat[Idle State]{\label{fig:frac_long_tcp_cam_id}\includegraphics[width=2.4 in ]{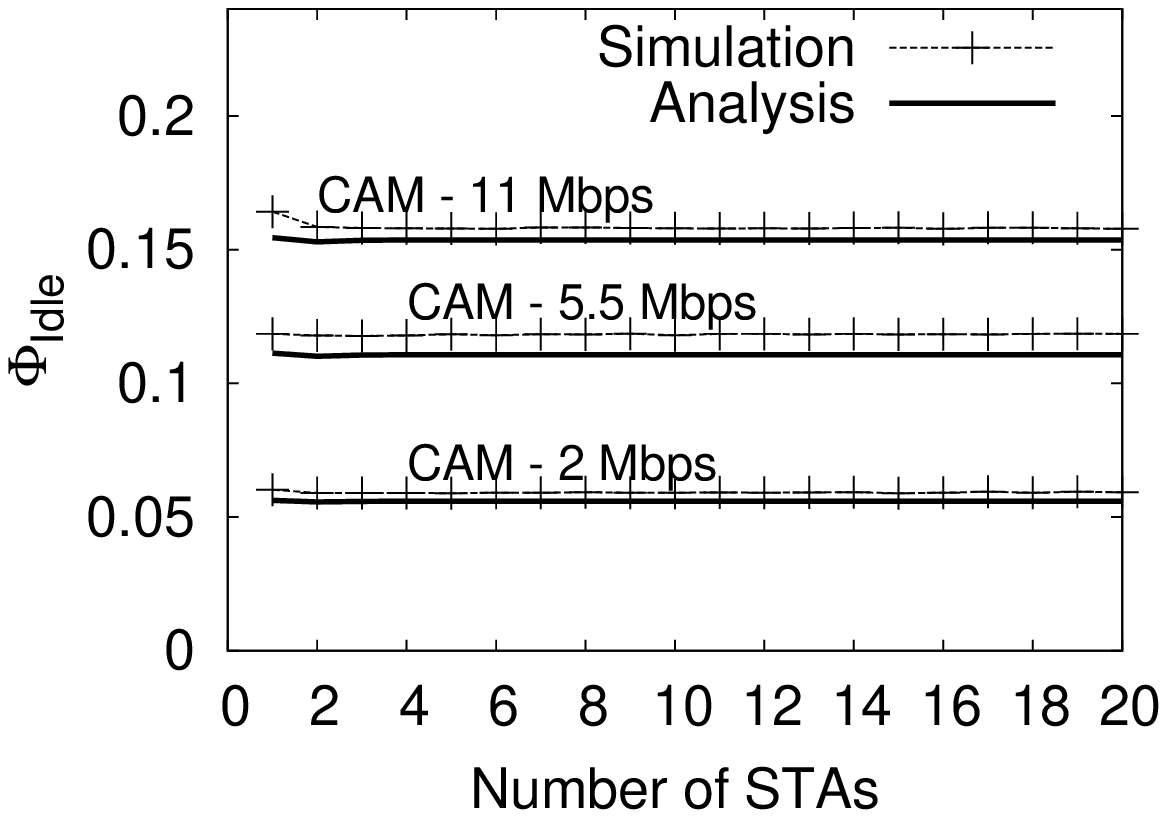}}
  \subfloat[Transmitting State]{\label{fig:frac_long_tcp_cam_tx}\includegraphics[width=2.4 in ]{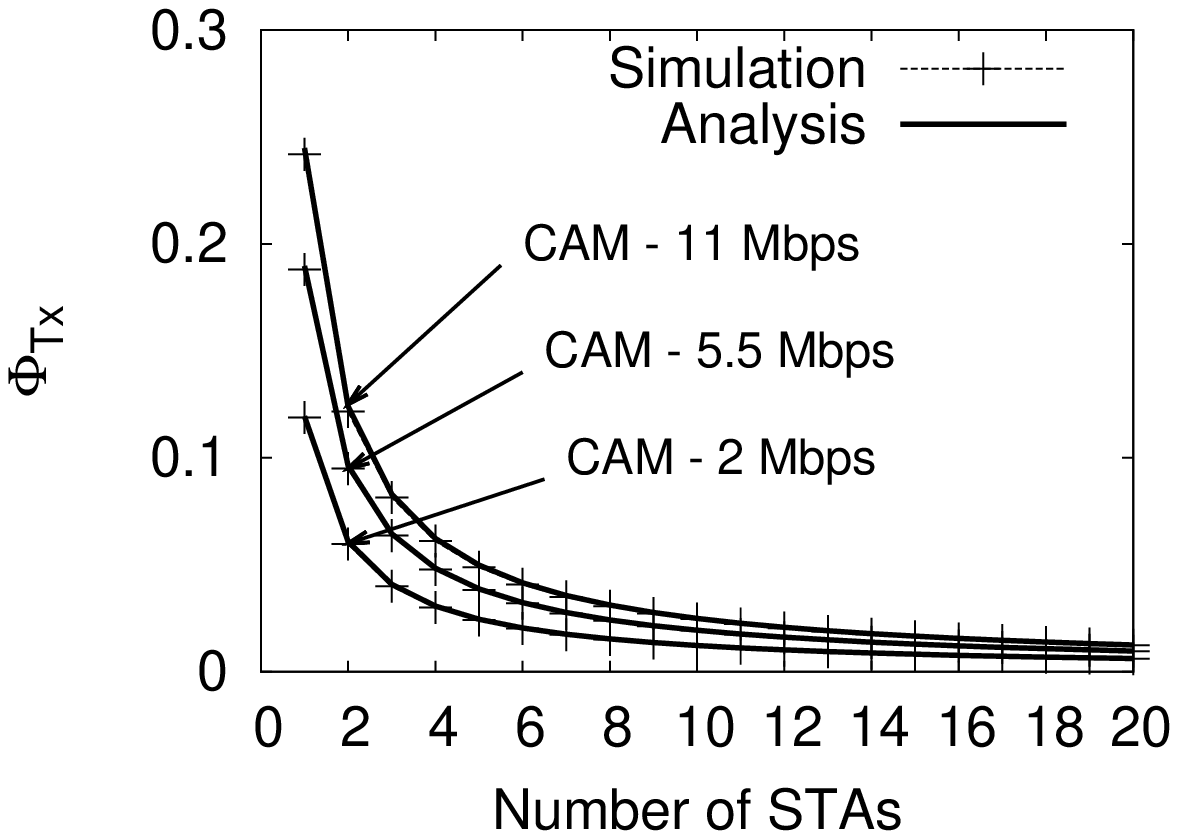}}
\subfloat[Receive \& Listen State]{\label{fig:frac_long_tcp_cam_ls}\includegraphics[width=2.4 in]{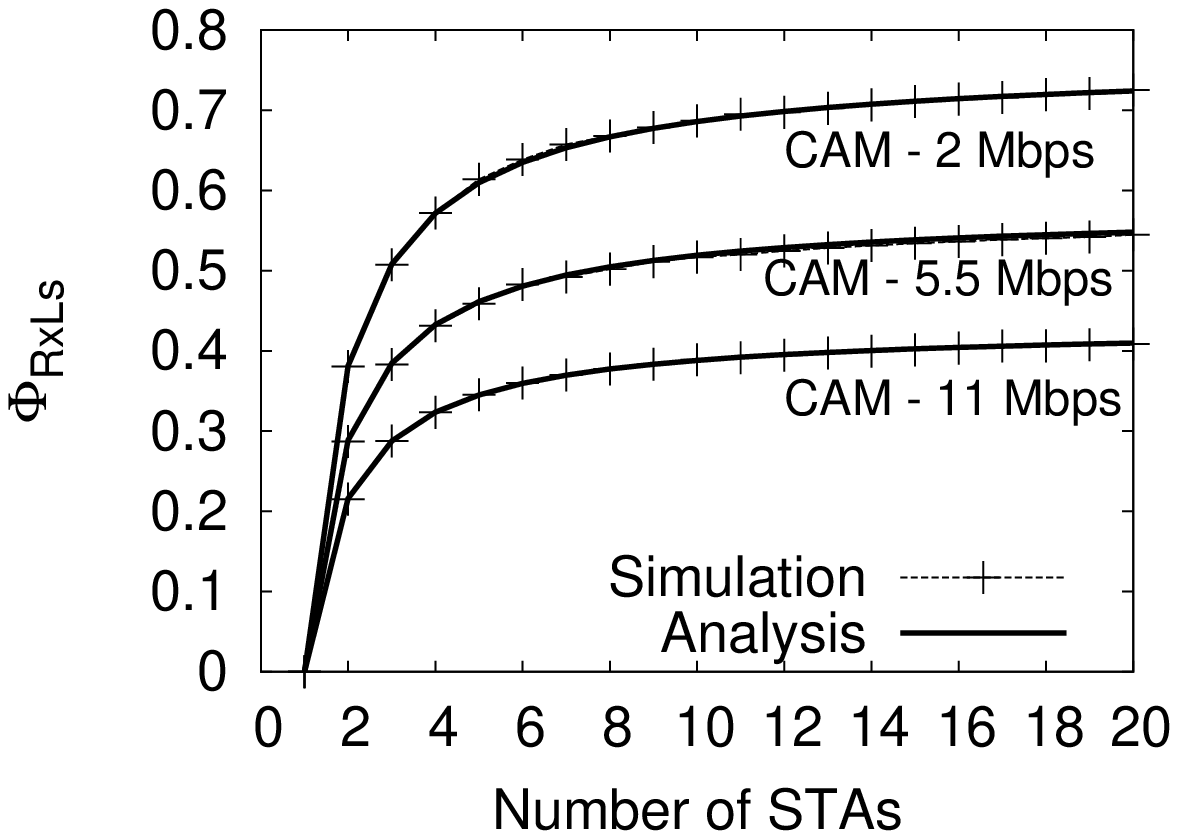}}\\
\subfloat[Receive \& Decode State ]{\label{fig:frac_long_tcp_cam_rxd}\includegraphics[width=2.4 in ]{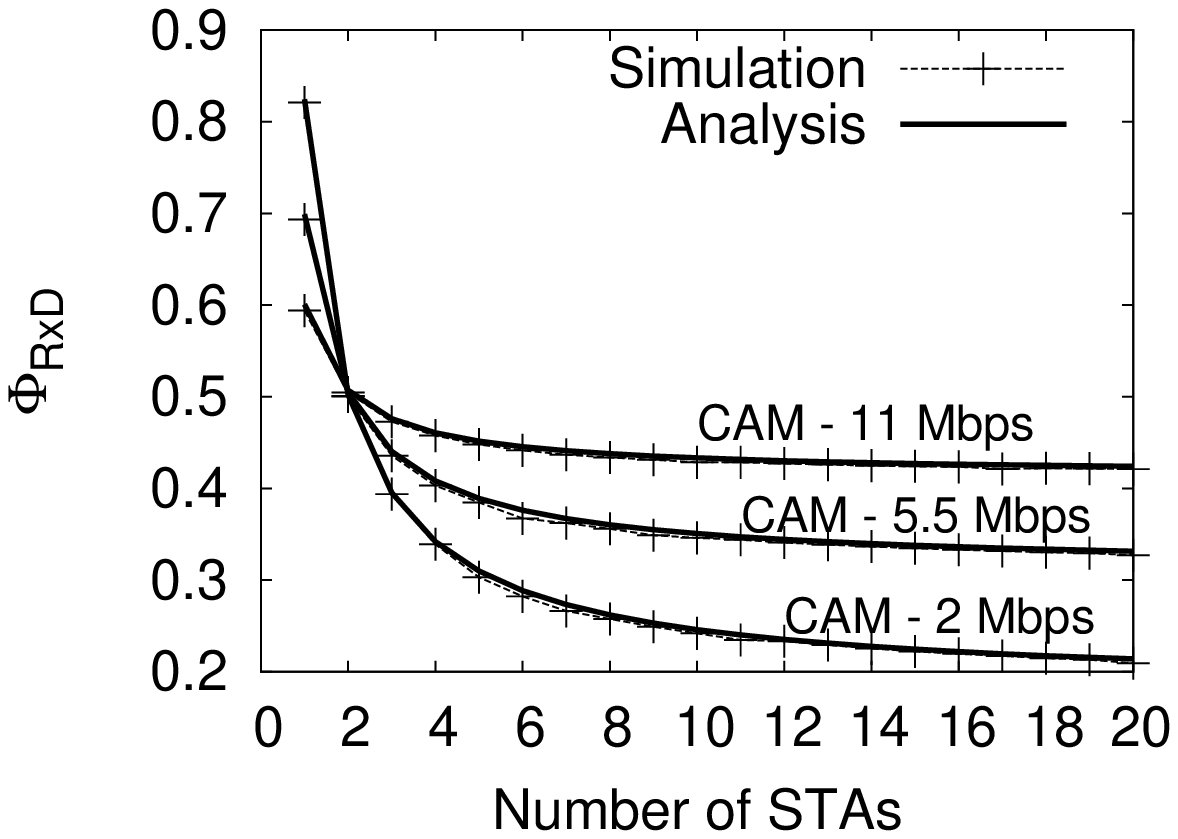}} 
\subfloat[Average Current (in mA) ]{\label{fig:avg_curr_long_tcp_cam}\includegraphics[width=2.4 in]{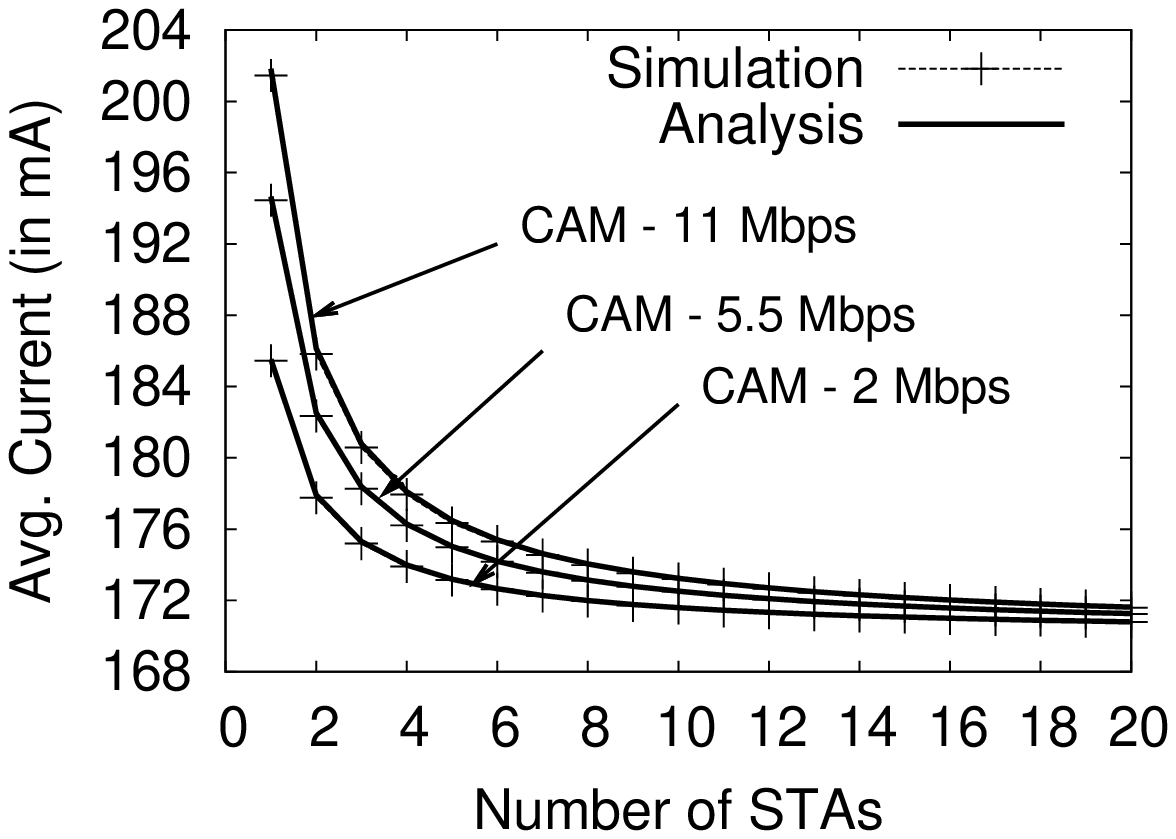}}
\subfloat[Efficiency (Mb / Coulomb) ]{\label{fig:eff_long_tcp_cam}\includegraphics[width=2.4 in]{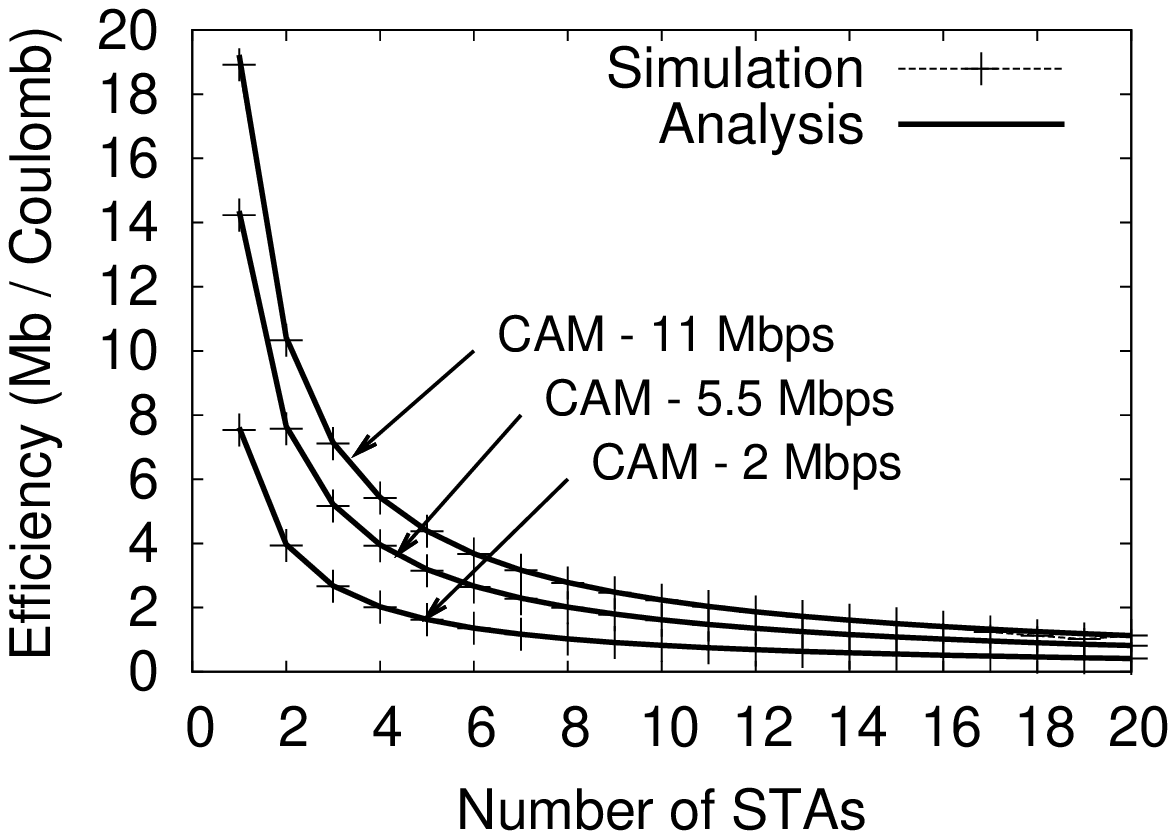}} 
  \caption{Continuous Active Mode}
  \label{fig:fracs_eff_cam_long}
\end{figure*}
\begin{figure*}
  \centering
\subfloat[Idle State]{\label{fig:frac_long_tcp_psm_id}\includegraphics[width=2.4 in ]{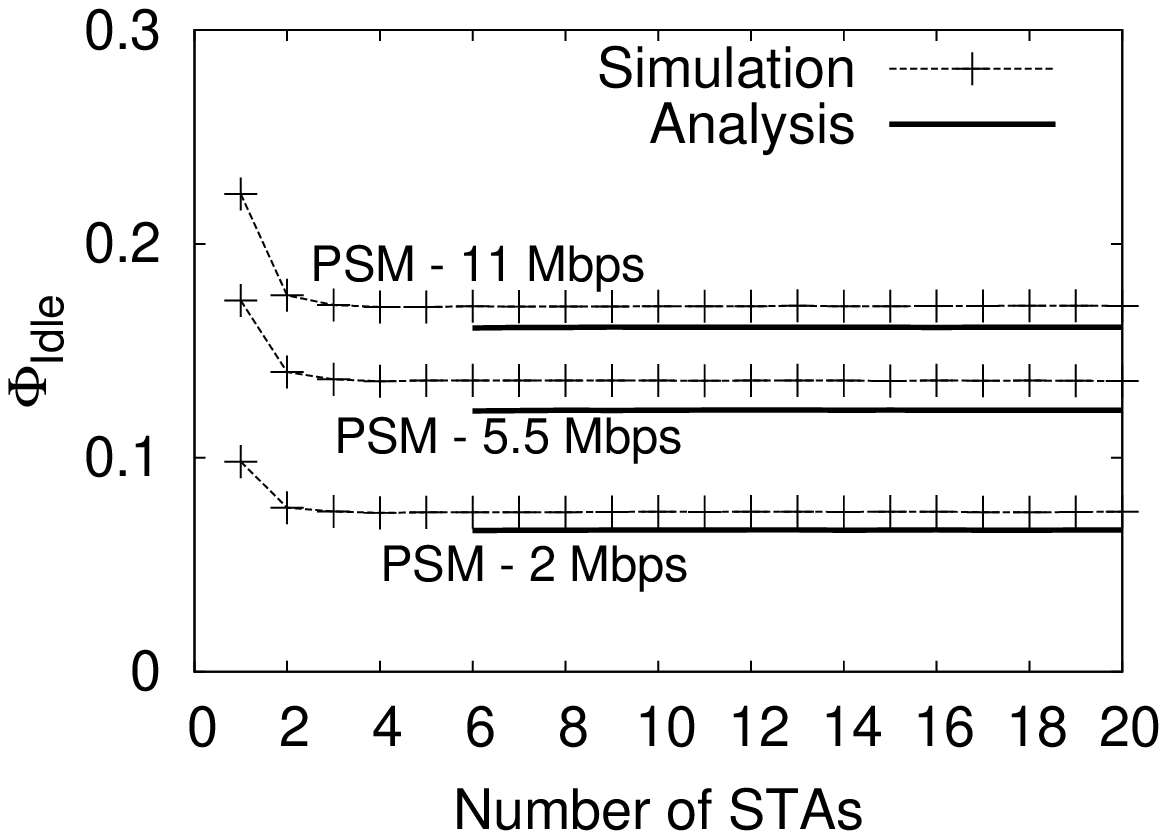}}
  \subfloat[Transmitting State]{\label{fig:frac_long_tcp_psm_tx}\includegraphics[width=2.4 in ]{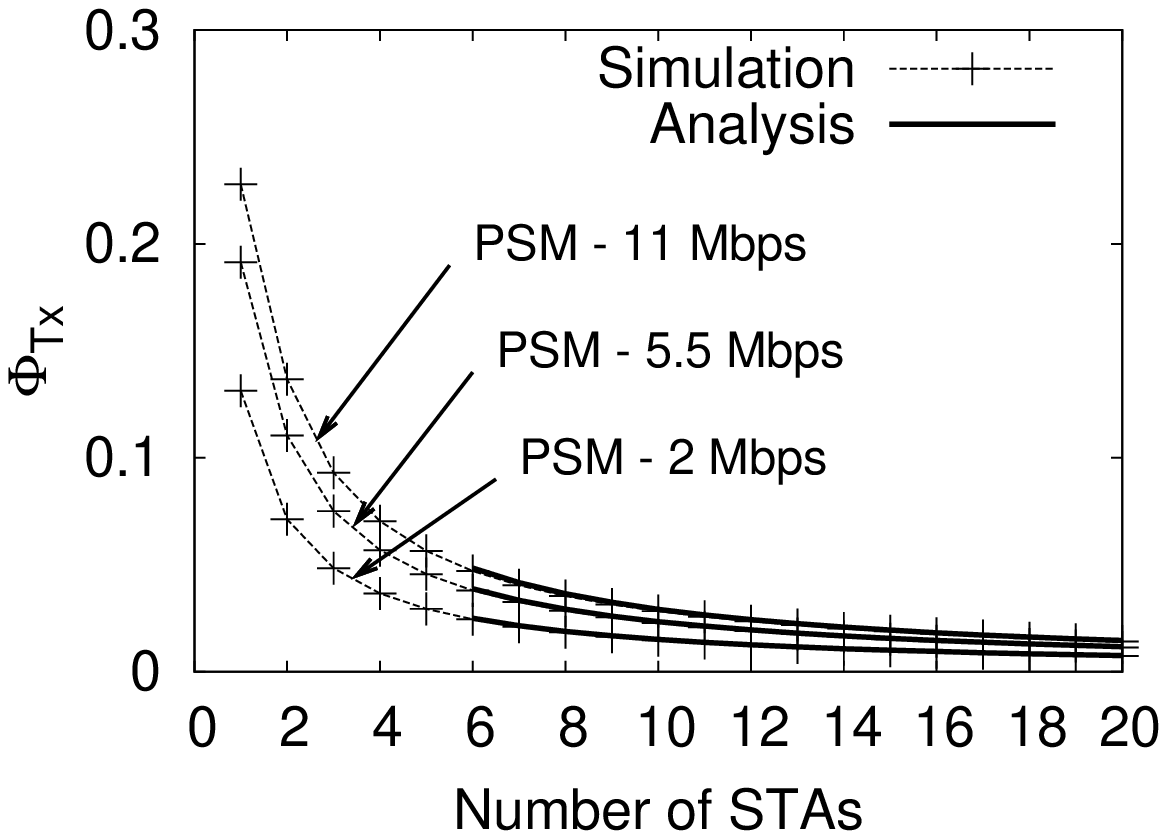}}
\subfloat[Receive \& Listen State]{\label{fig:frac_long_tcp_psm_ls}\includegraphics[width=2.4 in]{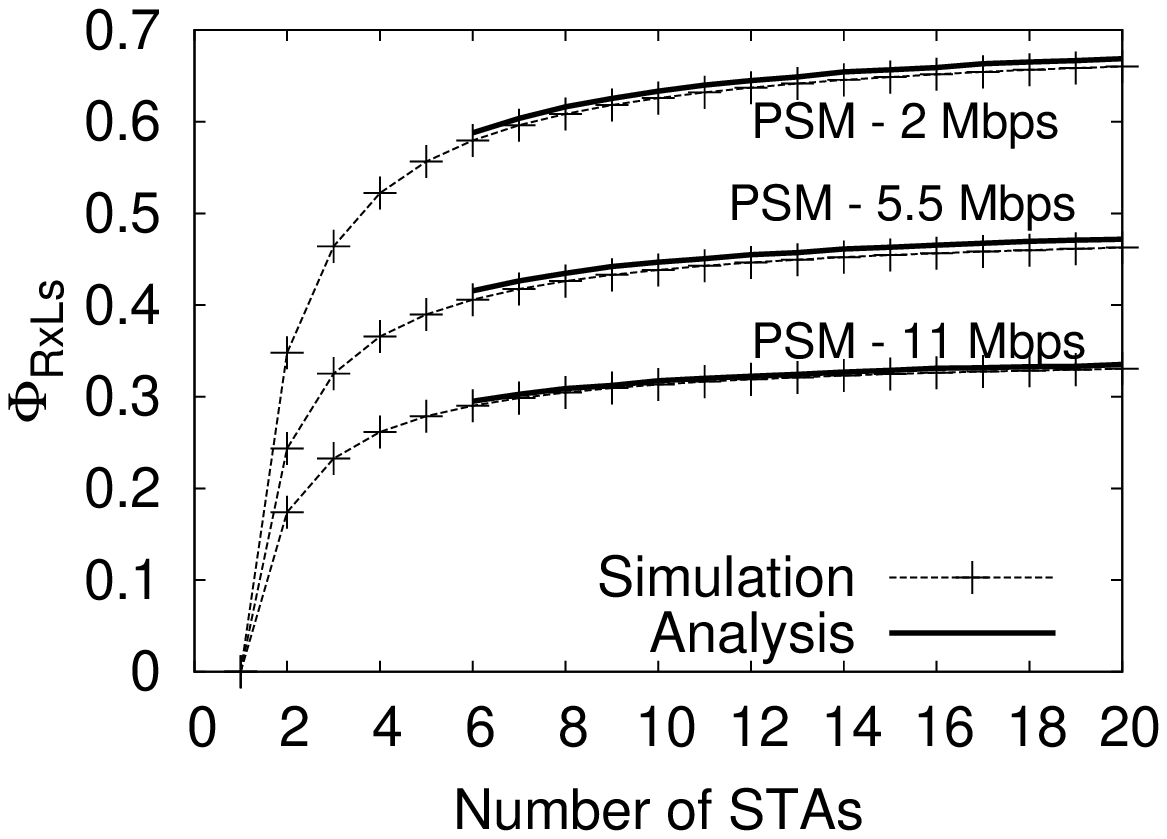}}\\
\subfloat[Receive \& Decode State ]{\label{fig:frac_long_tcp_psm_rxd}\includegraphics[width=2.4 in ]{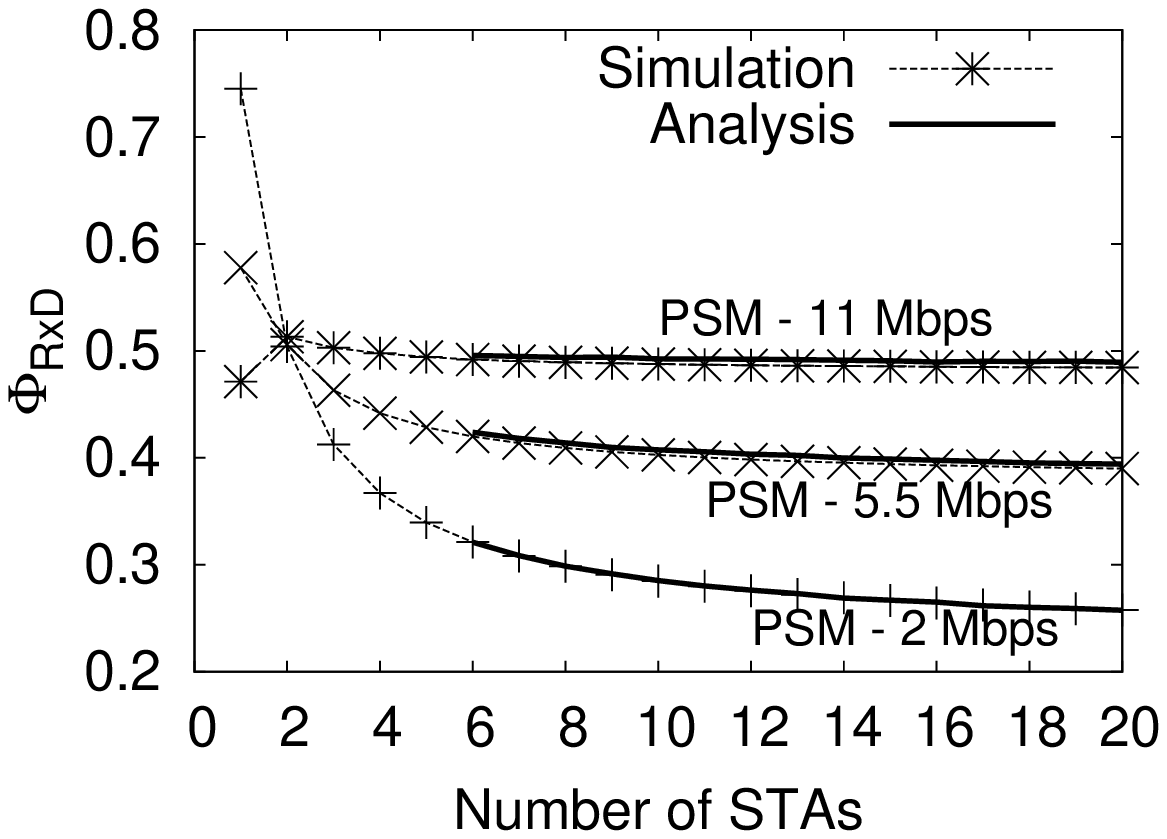}}
\subfloat[Average Current (in mA) ]{\label{fig:avg_curr_long_tcp_psm}\includegraphics[width=2.4 in]{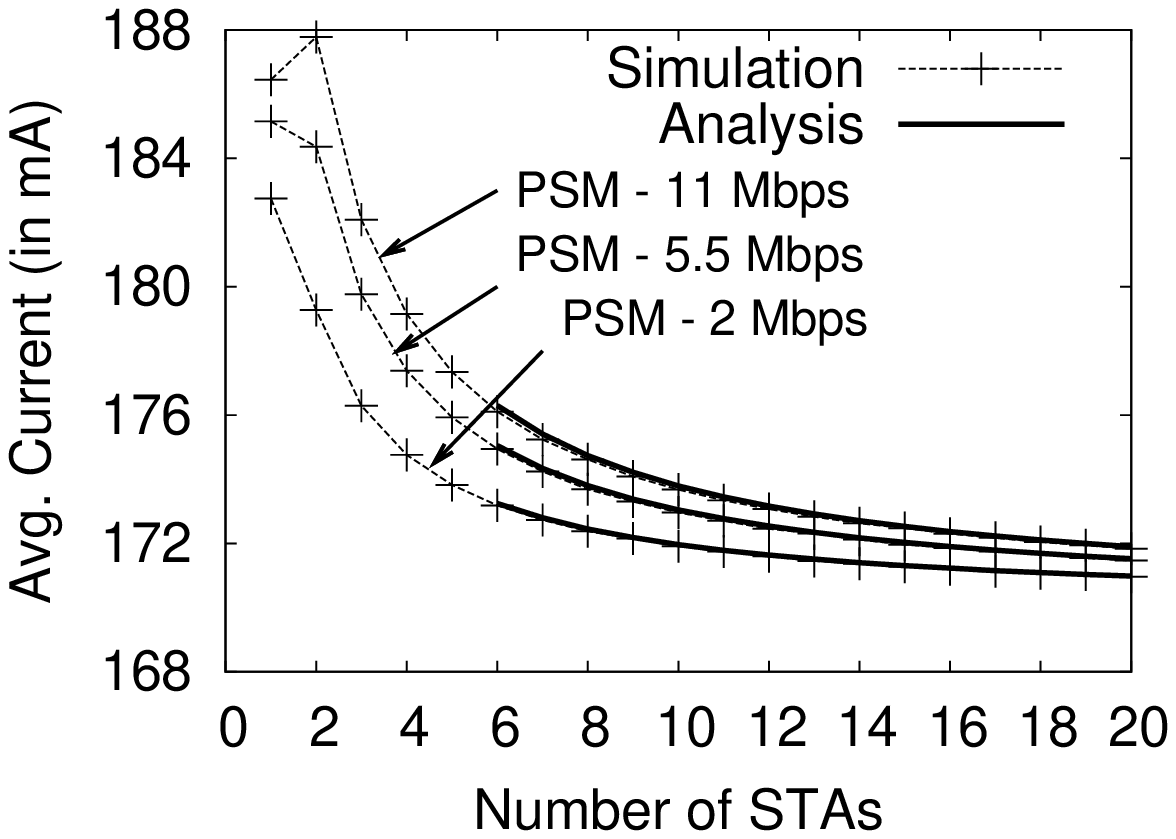}}
\subfloat[Efficiency (Mb / Coulomb) ]{\label{fig:eff_long_tcp_psm}\includegraphics[width=2.4 in]{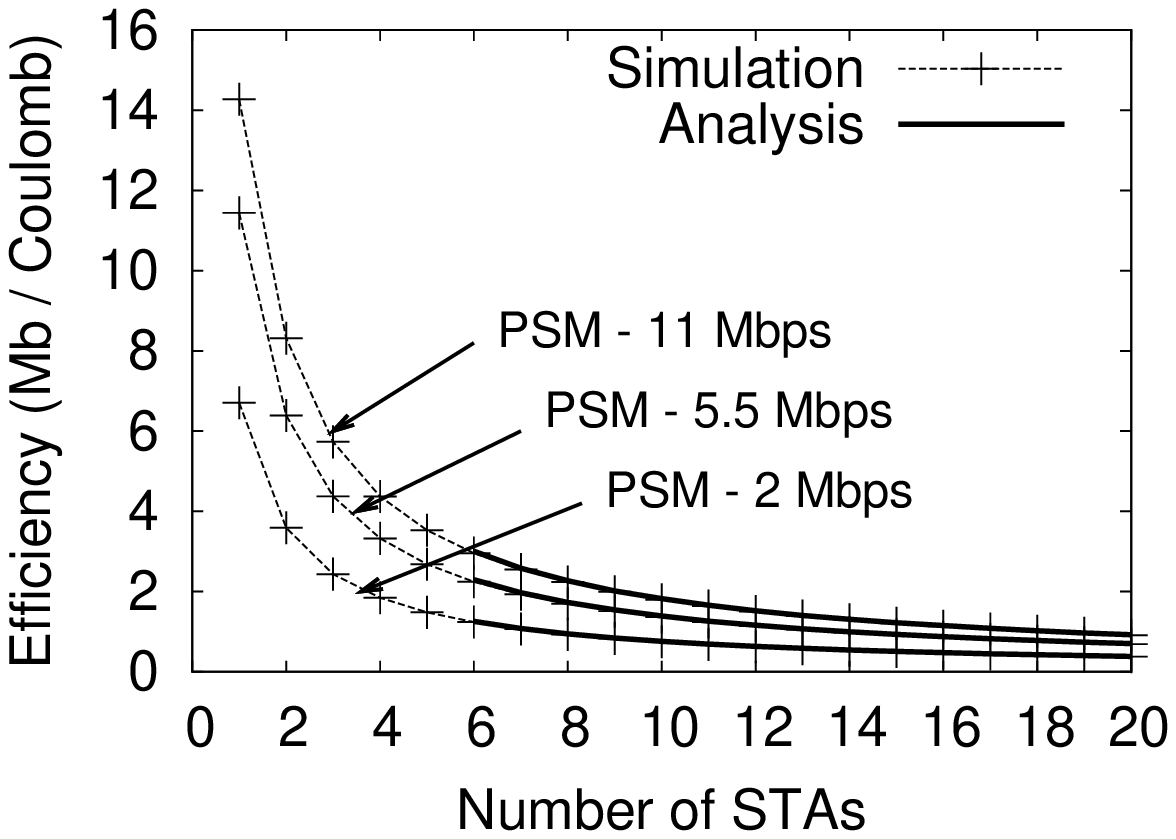}}
  \caption{Power Save Mode}
  \label{fig:fracs_eff_psm_long}
\end{figure*}
% \begin{figure}
%   \centering
%   \subfloat[Idle State]{\label{fig:frac_long_tcp_psm_id}\includegraphics[width=1.7 in ]{./eps/plots_long_tcp_psm/fracs_psm/frac_id.eps}}
%   \subfloat[Transmitting State]{\label{fig:frac_long_tcp_psm_tx}\includegraphics[width=1.7 in ]{./eps/plots_long_tcp_psm/fracs_psm/frac_tx.eps}}\\
% \subfloat[Listen State]{\label{fig:frac_long_tcp_psm_ls}\includegraphics[width=1.7 in]{./eps/plots_long_tcp_psm/fracs_psm/frac_ls.eps}}
% \subfloat[Receive \& decode State ]{\label{fig:frac_long_tcp_psm_rxd}\includegraphics[width=1.7 in ]{./eps/plots_long_tcp_psm/fracs_psm/frac_rxd.eps}}\\
% \subfloat[Average Current (in mA) ]{\label{fig:avg_curr_long_tcp_psm}\includegraphics[width=1.7 in]{./eps/plots_long_tcp_psm/metrics/avg_current.eps}}
% \subfloat[Efficiency (Mb / 1000 mA - s) ]{\label{fig:eff_long_tcp_psm}\includegraphics[width=1.7 in]{./eps/plots_long_tcp_psm/metrics/efficiency.eps}}\\
%   \caption{Fraction of time system remain various states}
%   \label{fig:fracs_psm_long}
% \end{figure}
\begin{figure}
   \centering
  \includegraphics[width
 = 3 in]{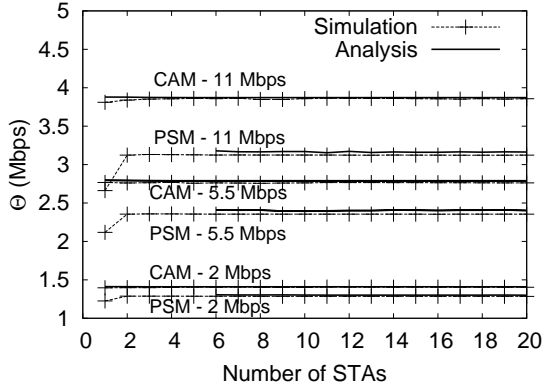}
 \caption{Aggregate Throughput - (CAM \& PSM)}
   \label{fig:throughput_long_tcp_cam_psm}
 \end{figure}
\subsection{Simulation Results -- Long Files} \label{sec:sim_res_long_files}
Simulation results are obtained using ns-2.33 and the various parameters used are taken from the 802.11b standard (given in Table ~\ref{tab:param}). Data rate is taken as $2$ Mbps to transmit control frames. To transmit data frames and MAC Header, data rate is taken as $02$, $5.5$, $11$ Mbps. The TCP packet size is taken as 1500B and the RTS threshold taken as 300B, which means that the TCP ACK is sent through basic access and the data packet is sent by RTS/CTS scheme. The values of current in different states of radio is taken from the specifications of the Intel PRO/Wireless 2011 card~\cite{tech_spec:intel_card_current}. Comparison of the throughput  obtained in CAM and PSM is shown in Fig. \ref{fig:throughput_long_tcp_cam_psm}. It can be seen that the aggregate throughput obtained by the STAs in PSM is less than that in CAM. The reason for this is the overhead of extra PS-POLL in case of PSM. Throughput in CAM is obtained using the model developed in Section ~\ref{sec:long_files_CAM}, and its analysis is shown in \cite{arxiv:our_paper_ana_mod}. 
%Similar model to evaluate the CAM throughput is developed in \cite{base:harsha_kuriakose} and \cite{Base:bcg_tcp_throughput_06}. 
 %Although our model for PSM is developed for the $N>5$ STAs, but the throughput obtained is even same for smaller number of STAs.
 % Results for the single STA in PSM downloading a large file over TCP is shown in Tab.~\ref{tab:psm_single_long}. There is a slight mismatch between the analysis and the simulation values. As discussed earlier STA goes to sleep state when there are no packets at the AP. When a single STA is downloading file then, then the partial TCP window remains at STA in the form of TCP ACKs and remaining at the AP in the form of data packets. If at any instant  whole TCP window comes in the form of TCP ACKs at the STA, it goes to sleep state, since the last packet it received must have More bit unset. STA remains in sleep state till the arrival of the next beacon frame, this results in lesser throughput and current values than the analytical values.
Figures~\ref{fig:fracs_eff_cam_long} and~\ref{fig:fracs_eff_psm_long} shows the comparison of analytical and simulation results for fraction of times, average current and efficiency. Efficiency is obtained by dividing the throughput (in Mbps) by average current (in mA), which is equivalent to data downloaded (in Mb) per Coulomb of charge drawn by an STA. 
Figures~\ref{fig:frac_long_tcp_cam_id} and~\ref{fig:frac_long_tcp_psm_id} shows the fraction of time an STA remains in the idle state for CAM and PSM respectively. It remains constant with number of STAs increasing. The time for which the channel remains in the idle state per data packet transmitted can be divided into three parts; 1) Time spent in backoff, this depends on the number of STAs contending, 2) Inter frame time, SIFS and DIFS, this remains constant, 3) Time spent in idling during collision, EIFS, this depends on the number of nodes contending. The throughput and the number of contending nodes \cite{base:harsha_kuriakose} do not change with number of STAs. So the time spent in decrementing backoff counter and the number of collisions per data packet transmitted also do not change. The interframe time, DIFS and SIFS, is constant for a data frame. It can be inferred that a data packet is associated with a constant idle time, irrespective of the number of STAs. Since the transmission and receive times of frames depend only on the data rate, so the fraction of time an STA stays in idle state remains constant.

The throughput share of a singe STA decreases, as number of STAs increases. It implies that number of STAs increases, an STA spends more time in receiving data frames for other STAs. The STAs stays in Receive \& Listen state (RxLs) state while it is receiving the data frames for other STAs. So the fraction of time an STA stays in RxLs state increases (Figs.~\ref{fig:frac_long_tcp_cam_ls} and ~\ref{fig:frac_long_tcp_psm_ls}) with increasing number of STAs. Also, because of this reason, the fraction of time any STA remains in Receive \& Decode (Figs.~\ref{fig:frac_long_tcp_cam_tx} and ~\ref{fig:frac_long_tcp_psm_tx}) and transmitting state decreases (Figs.~\ref{fig:frac_long_tcp_cam_rxd} and~\ref{fig:frac_long_tcp_psm_rxd}) with increasing number of STAs.

An STA transmits following frames per data frame it receives: CTS, MAC ACK and TCP ACK. Being control frames, CTS, MAC ACK are transmitted at $2$ Mbps, so the transmission time of CTS and MAC ACK does not change with data rates. Also, the transmission time of TCP ACK does not vary much with changing data rates because of its size ($98$ bytes). As the data rate increases, the throughout also increases, which implies that if we consider a time interval then with increasing data rates we can pack more data packets in it. When, number of transmissions of data frame increases, then number of transmissions of aforementioned frames also increases. Since the transmission time of these frames does not change with data rate, so the total transmission time increases, which lead to increase in the fraction of time an STA stays in transmission state with increasing data rate (Figs.~\ref{fig:frac_long_tcp_cam_tx} and ~\ref{fig:frac_long_tcp_psm_tx}).

% With changing data rates, inter frame times (SIFS, DIFS), time spent in decrementing backoff counter (excluding the duration when the channel is busy) and the time spent in idling during collision (EIFS), does not change. As the number back off slots decremented and number of collisions per transmitted data frame depends only on the number of contending nodes. 
Recalling, a data frame is associated with constant idle time. As data rate increases, in a given time interval, number of data packets transmitted also increases. So with increasing data rate the idle duration in the time interval increases, hence the fraction of time, an STA spends in idle state increases (Figs.~\ref{fig:frac_long_tcp_cam_id} and ~\ref{fig:frac_long_tcp_psm_id}).

Since the fraction of time an STA stays in idle and transmission state increases, so the fraction of time during which the STA stays in receive state (RxLs + RXD) decreases with increasing data rate. It could be obtained by adding the values shown by Figs. ~\ref{fig:frac_long_tcp_cam_ls} and ~\ref{fig:frac_long_tcp_cam_rxd} for CAM and Figs.~\ref{fig:frac_long_tcp_psm_ls} and ~\ref{fig:frac_long_tcp_psm_rxd} for PSM.

With the number of STAs increasing, the fraction of time an STA spends in transmitting state decreases and transmit current is more than the idle and receive current (Tab.~\ref{tab:param}), so with number of STAs increasing, the average current decreases (Fig.~\ref{fig:avg_curr_long_tcp_cam} for CAM and Fig.~\ref{fig:avg_curr_long_tcp_psm} for PSM) and converges to idle current for large number of STAs. Since the throughput of a single STA decreases as $1/N$ and the average current converges to a constant value, so the efficiency as defined above decreases as $1/N$. On comparing Fig.~\ref{fig:eff_long_tcp_psm} and Fig.~\ref{fig:eff_long_tcp_cam} it is clear that for the long file transfer case, CAM has higher efficiency than PSM, it is because of the overhead of PS-POLL in case of PSM.
%\begin{table}
%\renewcommand{\arraystretch}{1.4}
%\caption{For N = 1}
%\label{tab:for_N_1}
%\centering
%\begin{tabular}{|c|c|c|c|c|c|c|}
%\hline
%&\multicolumn{3}{c|}{$\Theta_1$ (Mbps)} &\multicolumn{3}{c|}{$J_1$ (mA)} \\ \hline
%            & 2     &   5.5  & 11      & 2        & 5.5     & 11 \\ \hline
%  Analysis  & $1.28$ & $2.33$ & $3.04$  & $213.7$   & $221.1$  & $226.1$  \\\hline
%Simulation  & $1.22$ & $2.11$ & $2.66$  & $208$     & $207.7$ &  $ 207.3$                          \\ \hline
%\end{tabular}
%\label{tab:psm_single_long}
%\end{table}
% \begin{table}
% \renewcommand{\arraystretch}{1.4}
% \caption{For N = 1}
% \label{tab:for_N_1}
% \centering
% \begin{tabular}{|c|c|c|c|c|c|c|}
% \hline
% &\multicolumn{3}{c|}{$\Theta_1$ (Mbps)} &\multicolumn{3}{c|}{$J_1$ (mA)} \\ \hline
%             & 2     &   5.5  & 11      & 2        & 5.5     & 11 \\ \hline
%   Analysis  & $1.28$ & $2.33$ & $3.04$  & $187.86$   & $197.37$  & $203.78$  \\\hline
% Simulation  & $1.22$ & $2.11$ & $2.66$  & $182.75$     & $185.15$ &  $ 186.45$                          \\ \hline
% \end{tabular}
% \label{tab:psm_single_long}
% \end{table}
\begin{table}
\renewcommand{\arraystretch}{1.3}
\caption{Parameters}
\begin{center}
\begin{tabular}{|l|p{1 cm}|l|l|}
\hline
\multicolumn{4}{|c|}{Parameter used} \\ \hline
Parameter              & Value            & Parameter      & Value     \\ \hline
 EIFS Time             & $364$ $\mu$s     & RTS Size       & 20 bytes   \\
 SIFS Time             & 10$\mu$s 	  & PS-POLL Size   & 20 bytes \\
 DIFS Time             & 50$\mu$s	  & CTS Size       & 14 bytes     \\
System Slot time       & 20$\mu$s         & MAC ACK Size   & 14 bytes      \\
PLCP Header time       & $144$ $\mu$s     & IP Header      & 20 bytes  \\
PHY Header time        & $48$ $\mu$s      & TCP data size  & $1500$ bytes \\
MAC Header Size        & $34$ bytes       &TCP Header Size    & 20 bytes    \\ 
$J_{Id}$ $J_{RxD}$, $J_{RxLs}$ & $170$ mA & $J_{Tx}$ & 300 mA                \\
 $J_{Sl}$                        & 10 mA    &          &                      \\
\hline
\end{tabular} \label{tab:param}
\end{center}
\end{table}
\section{Short Files} \label{sec:short_files}
Short file downloads and \emph{think times} between downloads is the typical behavior of a user browsing the Internet. We assume that all the files are part of a single TCP connection, which means that the TCP connection is established for the first file while for rest of the files, the same connection is used. For every file, an HTTP request packet is sent by STAs to initiate the transmission~\cite{rfc:http_persist}.

\begin{figure}
\begin{center}
 \scalebox{.6}{ \input{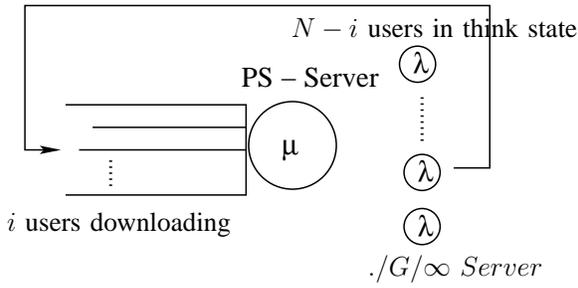}}
 \caption{Closed queueing network model for short file downloads. The service of file downloads at the WLAN is modeled by a processor sharing server; the bars behind the PS server represent the residual file sizes.}
\label{fig:ps_server}
\end{center}
\end{figure}

With the number of STAs increasing, the aggregate throughput of the AP does not change, as observed in the previous section, and this throughput is equally shared by all the STAs. Thus, the AP can be modeled as a Processor Sharing (PS) server and the think time can be modeled as the time spent at a $./G/\infty$ server. This is analogous to the Closed Queueing Network model in which there is a constant number of customers alternating between the Processor Sharing server (AP) and at a $./G/\infty$ server as shown in Fig.~\ref{fig:ps_server}.
Think time is considered to be exponentially distributed with mean $\frac{1}{\lambda}$ and file size distribution is taken as exponentially distributed with mean $L$. So the service time of a single file being downloaded alone is exponentially distributed with mean $\frac{1}{\mu} = \frac{L}{\Theta}$, where $\Theta$ is the aggregate throughput by STAs downloading large files, as obtained in the previous section.
For this scenario, we are interested in obtaining out two metrics: 
\begin{itemize}
 \item Average charge ($E[Q_f]$) per file -- It is defined as the, total charge drawn by all the STAs in a given interval divided by the total number of files downloaded by all them in the same interval.
\item Average sojourn time ($E[S]$) -- It is defined as the, total time taken by all the files downloaded in a given interval divided by the total number of files downloaded in the the same interval. Here, the time taken to download a files is taken as the time difference between the instant the STA starts contending for the HTTP request packet and the instant it receives the last packet of the file.
\end{itemize}
% \begin{figure}[htp]
% \centering
% \includegraphics[totalheight= 4 cm,width= 6 cm]{./eps/PS_server.eps}
% \caption{Closed queueing network model for short file downloads. The service of file downloads at the WLAN is modeled by a processor sharing server; the bars behind the PS server represent the residual file sizes.}\label{fig:ps_server}
% \end{figure}
\subsection{All STAs in CAM} \label{sec:short_files_CAM}
\begin{figure}
\centering
\includegraphics[totalheight= 1.6 cm]{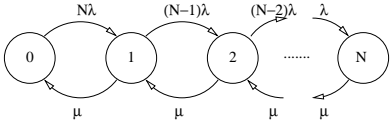}
\caption{Transition rate diagram of $X(t)$}\label{fig:ratetransition}
\end{figure}

If $X(t)$ is the number of ongoing short files transfers at $t$, then the number of STAs in the think state at $t$ will be $N-X(t)$.  $X(t)$ is a CTMC, because service time and think times are exponentially distributed. Fig.~\ref{fig:ratetransition} shows the transition rate diagram of $X(t)$.
\subsubsection{Expected charge drawn by a STA per short file downloaded ($E[Q_f]$) }
Define $J_{k,a}$ (derived in Section~\ref{sec:long_files_CAM}) as the average current drawn by $k$ STAs when they are downloading long files and let $J_{k,p}$ be the average current drawn by a STA listening (not doing any activity) to the traffic of $k$ STAs downloading long files (derived in Appendix ~\ref{appendix:passive_current}).
Let $I_{\{X(t)=k\}}$ be the indicator function indicating $k$ STAs active at any instant $t$. Let $Q_j(t)$ be the total charge drawn, and $n_j(t)$ be the number of downloads completed by STA $j$ in time interval ($0$, $t$). Our aim is to evaluate the average charge drawn by STAs per file, which is given by the following equation:
\begin{align}
%\begin{gather}
  E[Q_f]=\lim_{t \to \infty} \frac{\sum_{j=1}^N Q_j(t)}{\sum_{j=1}^N n_j(t)} = \lim_{t \to \infty} \frac{\frac{\sum_{j=1}^N Q_j(t)}{t}}{\frac{\sum_{j=1}^N n_j(t)}{t}}
\label{eqn:cam_short_charge0}
%\end{gather}
\end{align}
% The above equation can also be written as follows:
% \begin{equation}
% %\begin{gather}
%   E[Q_f]=  \lim_{t \to \infty} \frac{\frac{\sum_{j=1}^N Q_j(t)}{t}}{\frac{\sum_{j=1}^N n_j(t)}{t}}
% %\end{gather}
% \label{eqn:cam_short_charge1}
% \end{equation}
Note that if the limit (at $t\to \infty$) in the numerator and the denominator exist then these are, respectively, the rate of consumption of charge in all the STAs, and the total rate of transfer of short files (over all the STAs). Now, if there are exactly $k$ STAs active and downloading throughout the interval $(0,t)$ and $N-k$ in think state during that duration. Then the following expression gives the total charge drawn by all the STAs in the time interval $(0,t)$
\begin{align}
\int_0^t [kJ_{k,a} + (N-k)J_{k,p}]I_{\{X(t)=k\}} dt
\label{eqn:avg_chrg_cam_shrt}
\end{align}
On summing the above expression over all $k$, we get the sum of all the charge drawn by all the STAs, in the time interval $0$ to $t$. After summing the above expression and then substituting it in the Eqn.~\ref{eqn:cam_short_charge0}, we get the following equation:
% \begin{equation}
% \sum_{j=1}^N Q_j(t) = \sum_{k=0}^N\int_0^t [kJ_{k,a} + (N-k)J_{k,p}] I_{\{X(t)=k\}} dt
% \label{eqn:totcharge}
% \end{equation}
% Substituting the above equation in \ref{eqn:cam_short_charge1}, we get the following:
\begin{align}
\begin{split}
  E[Q_f]
 &=\lim_{t \to \infty} \frac{\frac{1}{t}\sum_{k=0}^N\int_0^t [kJ_{k,a} + (N-k)J_{k,p}] I_{\{X(t)=k\}} dt}{\frac{\sum_{j=0}^N n_j(t)}{t}} \\
&= \frac{\sum_{k=0}^N \pi_k [kJ_{k,a} + (N-k)J_{k,p}] }{\sum_{k=0}^{N}\pi_k(N-k)\lambda}\\
\end{split}
\label{eqn:cam_short_charge2}
\end{align}
where, $\pi_k$ is the stationary probability of $k$ STAs downloading files and $N-k$ STAs in think state.
\subsubsection{Expected Sojourn Time E[S]}
Similarly, the expression for expected sojourn time can be written as follows:
\begin{align}
%\begin{split}
E[S] = \frac{\sum_{k=0}^N k \pi_k}{\sum_{k=0}^{N}\pi_k(N-k)\lambda}
%\end{split}
\end{align}
%\begin{equation}
%\begin{split}
%\gamma= &\frac{\sum_{k=0}^N %\frac{u_k}{u_j}\frac{1}{a_j}(N-k)\lambda}{\sum_{k=0}^N \frac{u_k}{u_j} %\frac{1}{a_k}}
%\\=& \frac{\sum_{k=0}^N u_k\frac{1}{a_j}(N-k)\lambda}{\sum_{k=0}^N u_k %\frac{1}{a_k}}
%\label{eqn:Ueqaution}
%\end{split}
%\end{equation}
%where,
%$u_k$ is the solution of the  equation $u=uP$, and $P$ is the transition %probability matrix, of the EMC.
%Equation \ref{eqn:Ueqaution} can be written as follows,
% \subsubsection{Calculation of $\pi_k$}
% Assuming $\lambda>0$ and $\mu>0$, the CTMC is irreducible, and since there are finite states, it is positive recurrent. From the rate transition diagram shown in Figure~\ref{fig:ratetransition}, the following equations can be written:
% 
% \begin{equation}
% \begin{split}
% \pi_0 (N\lambda)=\pi_1 \mu \\
% \pi_1 (N-1)\lambda=\pi_2 \mu \
% \end{split}
% \end{equation}
% \\
% So $\pi_k$ can be expressed as,
% 
% \begin{equation}
% \pi_k=\left(\frac{\lambda}{\mu}\right)^k\frac{N!}{(N-k)!}\pi_0
% \end{equation}
% 
% As $\sum_{k=0}^N \pi_k = 1$, $\pi_k$ can be written as,
% \begin{equation}
% \pi_k=\frac{\left(\frac{\lambda}{\mu}\right)^k\frac{N!}{(N-k)!}}{{\sum_{k=0}^N\left(\frac{\lambda}{\mu}\right)^k\frac{N!}{(N-k)!}}}
% \end{equation} 
\subsection{All STAs in PSM} \label{sec:short_files_PSM}
In this scenario, STAs are in PSM, so when the user is in the think state, the STA goes to sleep state. When the user requests a file, the STA wakes up and sends a HTTP request packet and then again goes back to sleep. Since we are assuming the server to be local to the LAN, so the packets from the server in response to the request, arrives immediately at the AP. This information is sent to the STA in the next beacon frame. This means that the STA starts getting service at the beginning of the next beacon interval. After this the STA is assumed to remain in awake state until the whole file is downloaded. The interaction between the TCP slow start and the PSM~\cite{Experimental:ronny_bsd_05}, can be ignored in our case. It is a reasonable approximation because, RTT between the AP and the TCP server is negligible in our case, so the data packet arrive immediately in response to the TCP ACK, due to this the STA does not got to sleep state. Further, here we consider the file downloads in the presence of download type traffic to other STAs, this decreases the net throughput to a single STA. Hence, sojourn time of the file increases, so the time spent in slow start becomes less dominant.

Let $X(t)$ denote the number of STAs in the download state at time $t$, and $X_k$, $k \geq 0$ the value of $X(t)$ embedded at the beacon instants. Since the file sizes and think times are taken to be exponentially distributed, so the process $X_k$ is a DTMC. The transitions of the Markov chain are governed by the number of files completing transfer and the number of users completing their think times in the beacon interval. To make the calculation of transition probabilities simple, we assume that the 
users who complete their downloads starts their think times from the next beacon interval, so that the number of users that complete their think times in a beacon interval do not depend on the number of users who complete their transfers in the same beacon interval. This assumption is justified since the beacon interval is generally $100$ ms and the think time is generally of the order of seconds, hence the probability of a user completing its think time within one beacon interval is very small.
\subsubsection{Transition probabilities of the Markov Chain}
Let $N$ be the total number of STAs associated with the AP, and $b$ the duration of beacon interval.
Since we assume that the think times of STAs are exponentially distributed with mean $\frac{1}{\lambda}$, so the 
probability that user finishes his think time within interval of $b$ is $1 - e^{-\lambda b}$. If there are $i$ customers downloading files at the start of a beacon interval, then $N-i$ users are in think state, then the probability that $k$ users finish their think times within the beacon interval is $P_a(N,i,k) = \binom{N-i}{k}(1-e^{-\lambda b})^{k}(e^{-\lambda b})^{N-i-k}$. Let $q(i,m,b)$ be the probability that $m$ users complete their downloads out of $i$ active users within the time interval of $b$. This probability depends on the mean file length; we have assumed the files to be exponentially distributed with mean $\frac{1}{\mu} = \frac{L}{\Theta}$, where $\Theta$ is the throughput obtained in the previous section for the large file download case. Let us denote $p_{i,j}$  as the probability that there will be $j$ users downloading file at $X_{k+1}$, given that there were $i$ users downloading files at $X_k$. The transition probability ($p_{i,j}$) of $X_k$ can be written as follows:
% Consider $i < j$ and $m$ users completing downloads during the interval ($X_{k}, X_{k+1}$). If no user of these $m$ users complete his think time before $X_{k+1}$, then $j-i+m$ out of $N-i$ users must complete their think times in interval ($X_{k}, X_{k+1}$). As stated earlier, the probability that the users completing their downloads as well as their think times within the same beacon interval is negligible, so the transition probability ($p_{i,j}$) can be written as follows:
\begin{align}
\begin{split}
p_{i,j} &= \sum_{m=\max(0,i-j)}^{\min(i,N-j)}q(i,m,b) P_a(N,i,j,j-i+m) \\& \qquad \quad \qquad\qquad \qquad \qquad \mbox{$0<i,j\leq N$, $j=0$} \\
p_{0,j} &= \binom{N}{j}(1-e^{-\lambda b})^{j}(e^{-\lambda b})^{N-j}  \qquad \mbox{$0\leq j\leq N$} 
\end{split}
\label{eqn:transition_prob}
\end{align}
% \begin{equation*}
% p_{i,j}= \left\{ \begin{array}{l p{.5 cm}}
%  \sum_{m=\max(0,i-j)}^{\min(i,N-j)}q(i,m,b) \binom{N-i}{j-i+m}(1-e^{-\lambda b})^{j-i+m}(e^{-\lambda b})^{N-j-m} & \mbox{ $0<i\leq N$, $0\leq j\leq N$ }\\
%  \binom{N}{j}(1-e^{-\lambda b})^{j}(e^{-\lambda b})^{N-j} &\mbox{ $0\leq j\leq N$}
% \end{array} \right. 
% \end{equation*}
where, $q(i,m,b)$ is given by the following equation:
% \begin{equation}
% q(i,m,b) = \frac{\mu^m e^{-\mu b} b^{m}}{m!}
% \end{equation}
\begin{align}
q(i,m,b) = \left\{ \begin{array}{ll}
 \frac{\mu^m e^{-\mu b} b^{m}}{m!} & \mbox{$m < i$}\\
1 - \sum_{s=0}^{m-1}\frac{\mu^s e^{-\mu b} b^{s}}{s!} &\mbox{ $m = i$ }
\end{array} \right. 
\end{align}
% Let us define the following notations:\\
% $n_k(t)$ is the number of downloads completed by the user $k$ in $(0,t)$.\\ 
% % $S_k(t)$ is the total time spent by user $k$ in downloading $n_k(t)$ files. Note that, $S_k(t)$ does not include think times.\\ 
% % $Q_k(t)$ is the total charge drawn from the battery by customer $k$ during time interval $0$ to $t$. \\
% Recalling, $X(t)$ is the process representing the number of users downloading files at $t$, and $X_k$ is the number users downloading files at the $k^{th}$ beacon instant. \\
% % $I_{\{Y(t)=i\}}$ is the indicator function representing $i$ users downloading file at $t$. \\
% $\pi_i$ is the stationary probability of $X(t)$ \\
% $u_i$ is the stationary probability of $X_k$ \\
% $J_i$ is the average current drawn by the $i$ STAs, when $i$ are downloading files, and $N-i$ STAs are in think state.
\subsubsection{Calculation of Sojourn Time}
 Using Little's Theorem, following expression can be written for expected sojourn time:
\begin{align}
\begin{split}
E[S] =  \frac{\sum_{k=0}^{N} k \pi_k}{ \lim_{t \to \infty} \frac{1}{t} \sum_{j=1}^{N} n_j(t)}
\label{eqn:eq_s_3}
\end{split}
\end{align}
where, $n_j(t)$ is the number of downloads completed by the user $j$ in $(0,t)$, $\pi_k$ is the stationary probability of $X(t)$ 
The above expression only accounts for the time for which the STA stays in active state. It does not accounts for the time duration between the instant it sends the HTTP request and the next beacon instant. The expected value of this duration is $\frac{b}{2}$. By the definition, this duration is also included in the sojourn time. So the total sojourn time of the file is the sum of the above expression (Eqn.~\ref{eqn:eq_s_3}) and $\frac{b}{2}$.
\subsubsection{Calculation of average charge drawn per file}
In this scenario, STAs download a file and then go in Think state. During think time STAs stay in sleep state except when they wake up to listen to the Beacon Frames. As the Beacon frame is sent by contention so STA has to be awake for a some duration to be able to listen to it. Lets call this duration as $T_{Lb}$. The mean number of times STAs come to active state during think time is equal to the expected think time divided by the beacon interval ($\frac{1}{b\mu}$). Mean total duration for which STAs stays in active state during think time is $T_{Lb}(\frac{1}{b\mu})$. Using the Equations~\ref{eqn:cam_short_charge0} --~\ref{eqn:cam_short_charge2}, and taking the current drawn by the STAs in think state as $J_{Sl}$, following equation for the expected charge drawn per file can be written:
\begin{align}
\begin{split}
E[Q_f] = &\frac{\sum_{k=0}^{N} [k J_k + (N-k)J_{Sl}] \pi_k}{ \lim_{t \to \infty} \frac{1}{t} \sum_{j=1}^{N} n_j(t)} \\&+ J_{Id}T_{Lb}\left(\frac{1}{b\mu}\right) - J_{Sl}\left[\frac{1}{\lambda} -\left(\frac{1}{b\mu}\right) \right]
\end{split}
\label{eqn:eq_f_3}
\end{align}
where, $J_k$ is the average current drawn by the $k$ STAs, which are downloading files. It is to be noted that we have not modeled long files transfer in PSM scenario for $2\leq N\leq5$, so to evauate $J_2$ to $J_5$, we just extended the model of PSM for $N>5$. 

\subsubsection{Calculation of rate of arrivals} \label{section:appendix_rt_arrivals_psm_shrt}
$\lim_{t \to \infty }\frac{1}{t}\sum_{j=0}^{N}n_j(t)$ is given by  following expression:

\begin{equation}
\begin{split}
\lim_{t \to \infty } \frac{1}{t} \sum_{j=0}^{N}n_j(t) &= \frac{\sum_{k=0}^{N} u_k  E_k[n^{(a)}]}{b} \\
&=\frac{\sum_{k=0}^{N} u_k \left[ \sum_{l=1}^{N-k}l p^{(a)}(l,k)\right]}{b} 
\end{split}
\end{equation}
Here,\\
$u_k$ is the stationary probability of the Markov chain for the transition probabilities given in the Eqn.~\ref{eqn:transition_prob}.\\
$p^a(l,k)$ is the probability of $l$ arrivals in time interval $b$ when there are $k$ customers in service and it is given by following expression

\begin{equation}
p^a(l,k) = \binom{N-k}{l} (1- e^{-\lambda b})^l(e^{-\lambda b})^{N-k-l}
\end{equation}

\subsubsection{Calculation of Stationary Probability} \label{section:appendix_stat_pi_psm_shrt}
$\pi_k$ is calculated by using theory of MRGP
\begin{equation}
\begin{split}
\pi_k & = \lim_{t \to \infty} \frac{1}{t} \int_0^{t} I_{\{Y(t)=k\}}\,dt    \\&= \frac{\sum_{j=0}^{N}u_j E_k^{(j)}[T]}{\sum_{j=0}^{N}u_j b}\\
\end{split}
\end{equation}
$E_k^{(j)}[T]$  is the expected time spent in state $k$, between two regeneration points and starting with number of customers in the system equal to $j$.  \\
Detailed derivation of $E_k^{(j)}[T]$ is given in Appendix~\ref{section:appendix_expec_Ekj}

% Appendix~\ref{:mean_cycle_length_long_psm_N>5}.
% $\lim_{t \to \infty }\frac{1}{t}\sum_{j=0}^{N}n_j(t)$ is given by  following expression:
% \begin{equation}
% \begin{split}
% \lim_{t \to \infty } \frac{1}{t} \sum_{j=0}^{N}n_j(t) &= \frac{\sum_{k=0}^{N} u_k  E_k[n^{(a)}]}{b} \\
% &=\frac{\sum_{k=0}^{N} u_k \left[ \sum_{l=1}^{N-k}l p^{(a)}(l,k)\right]}{b} 
% \end{split}
% \end{equation}
% where, $u_k$ is the stationary probability of $X_k$, whereas $\pi_k$ is the stationary probability of $X(t)$.\\
% $p^a(l,k)$ is the probability of $l$ new downloads initiation in time interval $b$ when there are $k$ ongoing downloads $(l\leq N-k)$ and it is given by following expression:
% 
% \begin{equation}
% p^a(l,k) = \binom{N-k}{l} (1- e^{-\lambda b})^l(e^{-\lambda b})^{N-k-l}
% \end{equation}
\subsection{Simulation Results - Short Files}
Simulation results are obtained using ns-2.33 and the other parameters are same as stated earlier in Section~\ref{sec:sim_res_long_files}. To generate HTTP traffic in ns, we used PACKMIME~\cite{simulator:packmime}. The file size is taken to be exponentially distributed with mean $400KB$, and the think is taken to be exponentially distributed with mean $5 Secs$. The beacon interval is taken as $100 ms$ and the time duration for which the STA come in CAM, to listen to beacon frame is taken as $5 ms$.

\begin{figure*}
\subfloat[Expected Soj. Time]{\label{fig:soj_time_short_tcp_psm_all}\includegraphics[width=2.2 in]{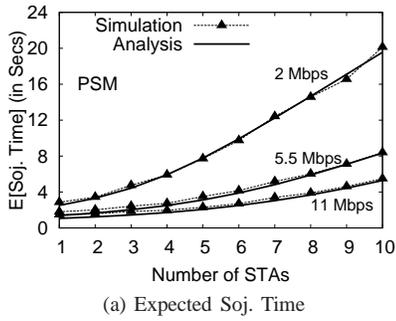}}
 \subfloat[No. of File Transfers / 100 C]{\label{fig:transac_psm_all}\includegraphics[width=2.2 in]{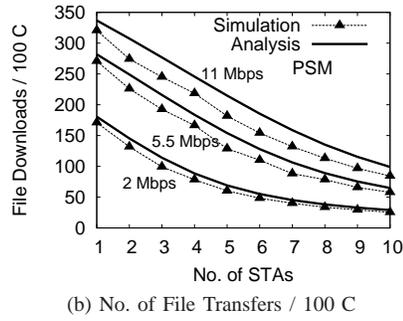}}
\subfloat[$\pi_k$ ($N = 8$)]{\label{fig:stat_prob_cam_N8_all}\includegraphics[width=2.2 in]{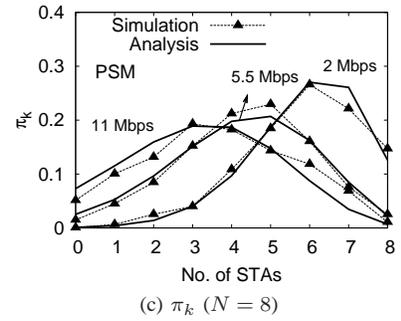}}
\caption{Power Save Mode: Short Files over TCP}
\end{figure*}

\begin{figure*}
\subfloat[Expected Soj. Time]{\label{fig:soj_time_short_tcp_cam_all}\includegraphics[width=2.2 in]{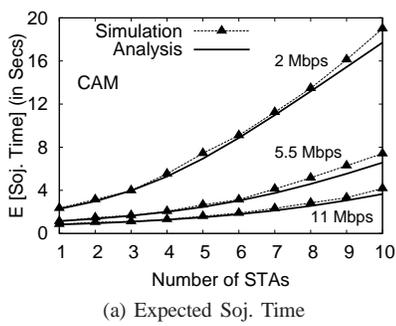}}
 \subfloat[No. of File Transfers / 100 C]{\label{fig:transac_cam_all}\includegraphics[width=2.2 in]{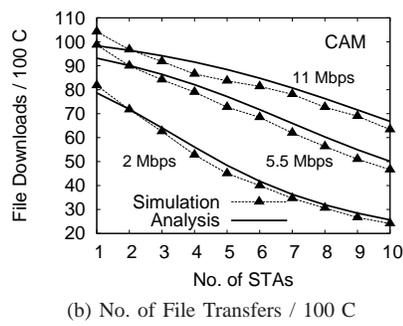}}
\subfloat[$\pi_k$ ($N = 8$)]{\label{fig:stat_prob_psm_N8_all}\includegraphics[width=2.2 in]{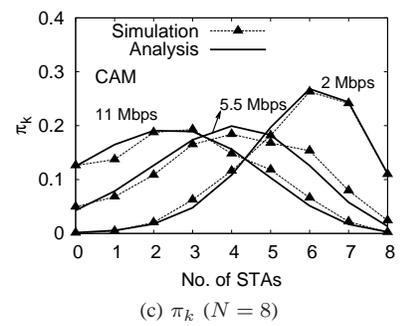}}
 \caption{Continuously Active Mode: Short Files over TCP}
\end{figure*}

Figures~\ref{fig:soj_time_short_tcp_cam_all} and \ref{fig:soj_time_short_tcp_psm_all} shows the comparison of sojourn time obtained using analysis and simulation, for PSM and CAM. It can be seen that the delay incurred in downloading file for CAM is slightly lesser than in PSM. This is due to lesser throughput achieved in PSM than in CAM. Figures~\ref{fig:transac_cam_all} and~\ref{fig:transac_psm_all} give the comparison of the simulation and the analytical values of the number of downloads that can be completed in a given battery capacity. Here, the battery capacity is taken in the form of maximum charge that can drawn from it. So the number of files that can be completed in a given battery capacity is obtained by dividing the battery capacity ($100 Coulomb$) by the expected charge drawn in downloading a file.

It is clear from Fig.~\ref{fig:transac_cam_all} and Fig.~\ref{fig:transac_psm_all} that Static PSM is more efficient than CAM. The reason behind this is that, the PSM STA goes to sleep state when it is not downloading anything; which is not the case with CAM. The PSM will even perform more better if the think time between the downloads increases, since then the CAM will be wasting more energy during idling. Further improvement in the PSM is possible by increasing the beacon interval, so that the STA does not have to wake up at every beacon instant, but it will increase the delay. It is to be noted, that with the number of STAs increasing the number of file downloads that can be completed in a given battery capacity decreases, because in this case while downloading its own file STA has to overhear the frame destined to other STAs also. Figs.~\ref{fig:stat_prob_cam_N8_all} and~\ref{fig:stat_prob_psm_N8_all} shows the stationary probability of $n$ station receiving service, when there are total of $N=8$ STAs associated with the AP. It is clear from the figures that there is considerable probability of more than one STA being active. Our future work will be focussing on this problem of decreasing efficiency with increasing number of STAs associated with the AP.
%#\vspace{-6 pt}
\section{Conclusion} \label{sec:conclusion}
In this paper, our contribution is two fold; firstly, we have modeled the energy consumption of TCP controlled large file transfers in CAM and in PSM, in the presence of download type TCP background traffic, which have been absent in the literature. Secondly, we modeled the energy consumption of TCP controlled short file transfers when all the STAs are in CAM and in PSM. We have seen that our analytical results matches quite well with that of the simulation results, which shows the correctness of our analysis. We have also shown that the PSM performs better than the CAM when the user remains inactive for some time in between the activity. However, if there is no inactivity then the performance of the PSM starts to degrade and performs worse than the CAM, as evident from the large file download case. In the future work, we will study the performance of PSM STA downloading short files, in the presence of CAM STAs carrying similar traffic. Further, we will study Adaptive PSM which will have features of both CAM and PSM; it does not have the extra overhead of PS-POLL and also can go to sleep state if user is not active for a certain time.

\onecolumn
\appendices
\section{PSM Short - Expected time spent in state $k$ } \label{section:appendix_expec_Ekj}

$E_k^{(j)}[T]$  is the expected time spent in state $k$, between two renewal points and starting with number of customers in the system equal to $j$ 

Let total number of customers in service be $i$ at $X_k$ and $u$ be any instant between $X_k$ and $X_{k+1}$.

Define $r(i,k,u)$ as the probability of $k$ customer departing from the system till time $u$ and the $k^{th}$ customer departs at instant $u$. $E_k^{(j)}[T]$  can be written as follows,

\begin{align}
\begin{split}
&E_k^{(j)}[T] = 0  \,\,\,\, j<k \\
&E_k^{(j)}[T] = b  \,\,\,\, j=k=0 \\
&E_k^{(j)}[T] = \int_{0}^{b} \mu u e^{-\mu u} du + b(e^{-\mu b}) \,\,\,\,  j= k \neq 0 \\
&E_k^{(j)}[T] = \int_{0}^{b}r(j,j-k,u)[\int_{u}^{b} (t-u)\mu e^{-\mu (t-u)} dt  +  (b-u)e^{-\mu (b-u)}]\,\, du \,\,\,\, j>k>0  \\
&E_k^{(j)}[T] = \int_{0}^{b}r(j,j-k,u)(b-u)\,\, du \,\,\,\,  j> k,\,\,\, k=0 
\label{eqn:expected_time_all_1}
\end{split}
\end{align}

Rearranging fourth case of equation \ref{eqn:expected_time_all_1}  we get the following equation,

\begin{equation}
\begin{split}
E_k^{(j)}[T] = \int_{0}^{b} \int_{u}^{b}  r(j,j-k,u)(t-u) \mu e^{-\mu (t-u)} dt\,du + \int_{0}^{b}  r(j,j-k,u)(b-u)e^{-\mu (b-u)} du \,\,\,\, for \,\,\,\, j> k >0 
\end{split} \label{eqn:expected_time_c5}
\end{equation} 
Changing the order of double integration in equation \ref{eqn:expected_time_c5} we get the following equation,
\begin{align}
\begin{split}
E_k^{(j)}[T] &= \int_{0}^{b} \int_{0}^{t}  r(j,j-k,u)(t-u) \mu e^{-\mu_k(t-u)} du\,dt + \int_{0}^{b}  r(j,j-k,u)(b-u)e^{-\mu (b-u)} du \\
E_k^{(j)}[T] &= \int_{0}^{b} (r(j,j-k) \star g)(t) dt + (r(j,j-k) \star g)(b)\,\,\,\, for \,\,\,\, j> k>0 \\
\end{split} \label{eqn:frth_case_2}
\end{align} 

Further fifth case of Eqn. \ref{eqn:expected_time_all_1} can be written as follows,
\begin{equation}
E_k^{(j)}[T] = (r(j,j-k)\star h)(b) du  \,\,\,\, for \,\,\,\, j> k,\,\,\, k=0  
\label{eqn:fifth_case_2}
\end{equation}

Where, \\
$g(t) = t\mu e^{-\mu t}$\\
$h(t) = t$\\
$r(j,j-k,u)$ can be expressed as follows,

\begin{align}
\begin{split}
r(j,j-k,u) &= f^{(j-k)}(u) \,\,\,\, for \,\,\,\,  j > k \\
r(j,j-k,u) &= e^{-\mu u} \,\,\,\, for \,\,\,\,  j = k 
\end{split}
\label{eqn:r_prob_1}
\end{align}
Taking the laplace of the Eqn. \ref{eqn:r_prob_1}, we get the following equation,
\begin{align}
\begin{split}
R(j,j-k,s) &=  \left(\frac{\mu}{\mu + s}\right)^{j-k}  \,\,\,\, for \,\,\,\,  j > k \\
R(j,j-k,s) &= \frac{1}{\mu + s}  \,\,\,\, for \,\,\,\,  j = k 
\end{split}
\label{eqn:r_prob_1_lap}
\end{align}
Taking the laplace of  \ref{eqn:frth_case_2}, and substituting Eqn \ref{eqn:r_prob_1_lap} in it, we get the following equation
\begin{equation}
\begin{split}
&E_k^{(j)}[T] =  \int_{0}^{b} (r(j,j-k) \star g)(t) dt + (r(j,j-k) \star g)(b)\,\,\,\, for \,\,\,\, j> k>0 \\
& L[E_k^{(j)}[T]] =  R(j,j-k,s) G(s)\frac{1}{s} + R(j,j-k,s) G(s)\,\,\,\, for \,\,\,\,  j>k>0 \\
&= \left(\frac{\mu}{\mu + s}\right)^{j-k} \frac{\mu}{(s + \mu)^2} \frac{1}{s} + \left(\frac{\mu}{\mu + s}\right)^{j-k} \frac{\mu}{(s + \mu)^2} \\
 &= \mu^{(j-k+1)}\left(\frac{1}{\mu + s}\right)^{j-k+2} \frac{1}{s} + \mu^{(j-k+1)}\left(\frac{1}{\mu + s}\right)^{j-k+2}   \\
& = \mu^{(j-k+1)} \left[\sum_{l=1}^{j-k+2}\frac{-1}{\mu^{j-k+3-l}(\mu + s)^l} + \frac{1}{s\mu^{j - k + 2}}\right] + \mu^{(j-k+1)}\left(\frac{1}{\mu + s}\right)^{j-k+2}  \\
\end{split}
\label{eqn:frth_case_3}
\end{equation}
Taking the inverse of Eqn. \ref{eqn:frth_case_3}, we get the following equation,
\begin{equation}
\begin{split}
 E_k^{(j)}[T] &= \mu^{(j-k+1)} \left[\sum_{l=1}^{j-k+2}\frac{-1 t^{l-1}e^{-\mu t}}{\mu^{j-k+3-l}(l-1)!} + \frac{1}{\mu^{j - k + 2}}\right] + \mu^{(j-k+1)} \frac{t^{j-k + 1}e^{-\mu t}}{(j-k + 1)!} 
\end{split}
\end{equation}
Taking the laplace of Eqn. \ref{eqn:fifth_case_2}, and substituting Eqn \ref{eqn:r_prob_1_lap} in it, we get the following equation
\begin{equation}
\begin{split}
E_k^{(j)}[T] &= (r(j,j-k)\star h)(b)\,du\,\,\,\, for \,\,\,\, j> k ,\,\, k=0 \\
L[E_k^{(j)}[T] ]&= R(j,j-k,s) H(s) \,\,\,\, for \,\,\,\, j> k ,\,\, k=0 \\
& = \left(\frac{\mu}{\mu + s}\right)^{j-k}  \frac{1}{s^2} \\
		& = \mu^{j-k}\left[\sum_{l = 1}^{j-k} \frac{j- k + 1 - l}{\mu^{j-k+2-l}(\mu + s)^l}  - \frac{j-k}{s\mu^{j-k+1}} + \frac{1}{s^2\mu^{j-k}}\right] 
\end{split}
\label{eqn:fifth_case}
\end{equation}
Taking the inverse of Eqn. \ref{eqn:fifth_case}, we get the following equation,
\begin{equation}
\begin{split}
&E_k^{(j)}[T]  = \mu^{j-k}\left[\sum_{l = 1}^{j-k} \frac{(j- k + 1 - l)e^{-\mu t}t^{l-1}}{(l-1)!\mu^{j-k+2-l}}  - \frac{(j-k)}{\mu^{j-k+1}} + \frac{t}{\mu^{j-k}}\right]
\end{split}
\end{equation}
Equation \ref{eqn:expected_time_all_1} can be written as follows
\begin{align}
\begin{split}
E_k^{(j)}[T] &= 0 \qquad j<k \\
E_k^{(j)}[T] &= b \qquad  j=k=0 \\
E_k^{(j)}[T] &= \int_{0}^{b} \mu u e^{-\mu u} du + b(e^{-\mu b})  \qquad j= k \neq 0 \\
E_k^{(j)}[T] &= \mu^{(j-k+1)} \left[\sum_{l=1}^{j-k+2}\frac{-1 t^{l-1}e^{-\mu b}}{\mu^{j-k+3-l}(l-1)!} + \frac{1}{\mu^{j - k + 2}}\right] + \mu^{(j-k+1)} \frac{t^{j-k + 1}e^{-\mu b}}{(j-k + 1)!} \qquad j> k>0 \\
E_k^{(j)}[T] &= \mu^{j-k}\left[\sum_{l = 1}^{j-k} \frac{e^{-\mu b}b^{l-1}}{(l-1)!\mu^{j-k+2-l}}  - \frac{(j-k)}{\mu^{j-k+1}} + \frac{b}{\mu^{j-k}}\right]  \qquad j>k,\,\,\, k=0 
\end{split}
\label{eqn:expected_time_all_2}
\end{align}
\section{Mean cycle length - PSM LONG (N$>$5)} \label{appendix:mean_cycle_length_long_psm_N>5}
Let the attempt probability for of a node when there are $n$ nodes saturated, obtained by fixed point analysis, be $\beta_n$, Following recursive equation can be written for $E_{i,j} = E_{i,j}[T_k]$,
\begin{align}
\begin{split}
 E_{i,j} = &P_{idle}^{(i,j)}[\delta + E_{(i,j)}] + P_{sR}^{(i,j)}[T_{sR}] + P_{sT}^{(i,j)}T_{sT} + P_{sPSPL}^{(i,j)}[T_{sP}] + P_{cT}^{(i,j)}[T_{cT} + E_{(i,j)}] \\& + P_{cP}^{(i,j)}[T_{cP} + E_{(i,j)}] + P_{cP-R}^{(i,j)}[T_{cP-R} + E_{(i,j)}] \\&+ P_{cT-R}^{(i,j)}[T_{cT-R} + E_{(i,j)}] + P_{cT-P}^{(i,j)}[T_{cT-P} + E_{(i,j)}] +  P_{cT-P-R}^{(i,j)}[T_{cT-P-R} + E_{(i,j)}]\\ \\
E_{i,j} = &\frac{P_{idle}^{(i,j)}[\delta] + P_{sR}^{(i,j)}T_{sR} + P_{sTACK}^{(i,j)}T_{sT}}{1 - P_{c}^{(i,j)} - P_{idle}^{(i,j)}}  + \frac{P_{sP}^{(i,j)}T_{sP} + P_{cT}^{(i,j)}T_{cT}}{1 - P_{c}^{(i,j)} - P_{idle}^{(i,j)}} 
+ \frac{P_{cP}^{(i,j)}T_{cP}  + P_{cP-R}^{(i,j)}T_{cP-R} + P_{cT-R}^{(i,j)}T_{cT-R}}{1 - P_{c}^{(i,j)} - P_{idle}^{(i,j)}}  \\&+ \frac{ P_{cT-P}^{(i,j)}T_{cT-P} + P_{cT-P-R}^{(i,j)}T_{cT-P-R} }{1 - P_{c}^{(i,j)} - P_{idle}^{(i,j)}}  
\end{split}
\end{align}

Notations used in the above equation are defined below  and they uses the 802.11 parameters defined in the Table~\ref{tab:param} 
\\
 \begin{tabular}{ p{2cm}  p{12cm}}
    $T_{sR}$ & It is the time required for transmitting one TCP data packet from AP = $T_{DIFS}$ + $T_{DATA}$ + $3T_{SIFS}$ + $T_{ACK}$ \\
    $T_{sT}$ & It is the time required for transmitting one TCP ACK packet = $T_{DIFS}$ + $T_{TACK}$ + $T_{SIFS}$ + $T_{ACK}$\\
    $T_{sP}$ & It is the time required for transmitting PS-POLL packet =  $T_{DIFS}$  + $T_{PSPL}$ + $T_{SIFS}$ + $T_{ACK}$\\
    $T_{cT}$ & It is the time spent in collision, when collision invlolves only TCP ACK packets = $T_{TACK}$ + $T_{EIFS}$\\
    $T_{cP}$ & It is the time spent in collision, when collision invlolves only PS POLL packets $T_{PSPL}$ + $T_{EIFS}$ \\
    $T_{cP-R}$ & It is the time spent in collision, when collision invlolves only PS POLL packets =  $max(T_{PSPL}, T_{RTS})$ + $T_{EIFS}$ \\
    $T_{cT-R}$ & It is the time spent in collision, when collision invlolves TCP ACK and RTS  = $max(T_{TACK}, T_{RTS})$ + $T_{EIFS}$ \\
$T_{cP-T}$ & It is the time spent in collision, when collision invlolves PS POLL and TCP ACK = $max(T_{TACK}, T_{PSPL})$ + $T_{EIFS}$ \\
$T_{cP-T}$ & It is the time spent in collision, when collision invlolves PS POLL and TCP ACK  = $max(T_{TACK}, T_{PSPL})$ + $T_{EIFS}$ 
\label{tab:psm_times_long_more_than_5}
  \end{tabular} 

\begin{tabular}{ p{2cm}  p{12cm}}
$T_{cP-T-R}$ & It is the time spent in collision, when collision invlolves PS-POLL, TCP ACK, and RTS = $max(T_{TACK}, T_{PSPL}, T_{RTS}) + T_{EIFS}$ 
\end{tabular} 
\\
Let's define the following:
\begin{equation}
\begin{split}
&r = i + j + 1 \,\,for \,\, (i,j) \neq (N,0) \\
&r = i + j  \,\,for \,\, (i,j) = (N,0) \\
&P_1(r,i) = 1 - (1 - \beta_{r})^{(i)} \\
&P_1(r,j) = 1 - (1 - \beta_{r})^{(j)} \\
&P_2(r,i) = 1 - (1 - \beta_{r})^{(i)} - i(1 - \beta_{r})^{(i-1)} \beta_{r}\\
&P_2(r,j) = 1 - (1 - \beta_{r})^{(j)} - j(1 - \beta_{r})^{(j-1)} \beta_{r}\\
\end{split}
\end{equation}

\begin{tabular}{ p{2 cm}  p{12cm}}
$P_{idle}^{(i,j)}$  		& It is the probability of a slot being idle = $(1 - \beta_{r})^{r}$ \\
$P_{sR}^{(i,j)}$ 		& It is the probability AP wins the contention 
				\begin{equation*}
				= \left\{ \begin{array}{ll}
            			\beta_{r}(1 - \beta_{r})^{r-1} & \mbox{ $(i,j)\neq (N,0)$ }\\
         			0 &\mbox{ $(i,j) = (N,0)$ }
			        \end{array} \right. 
				\end{equation*}\\
$P_{sT}^{(i,j)}$ 		& It is the probability STA with TCP ACK at HOL wins the contention 
				= $j\beta_{r}(1 - \beta_{r})^{(r-1)}$\\
$P_{sP}^{(i,j)} $		& It is the probability STA with PS-POLL at HOL wins the contention = $i\beta_{r}(1 - \beta_{r})^{r-1}$ \\
$P_{cP}^{(i,j)}$ 		& It is the probability that there is a collision between PS-POLL packets  = $P_2(r,i)(1 - \beta_{r})^{r-i}$ \\
$P_{cT}^{(i,j)}$ 		&It is the probability that there is a collision between TCP ACK packets  = $P_2(r,j)(1 - \beta_{r})^{r-j}$ \\
$P_{cT-P}^{(i,j)}$ 		&It is the probability that there is a collision between PS-POLL and TCP ACK packets  = $P_1(r,i)P_1(r,j)(1 - \beta_{r})^{(r-i-j)}$ \\
$P_{cT-R}^{(i,j)}$ 		& It is the probability that there is a collision between TCP ACK and RTS packets  
			\begin{equation*}
= \left\{ \begin{array}{ll}
            (1 - \beta_{r})^{(i)}P_1(r,j)\beta_{r} & \mbox{ $(i,j) \neq (N,0)$ }\\
         0 &\mbox{ $(i,j) = (N,0)$ }
			           		\end{array} \right. 
\end{equation*}\\

$P_{cP-R}^{(i,j)} $		&It is the probability that there is a collision between PS-POLL and RTSpackets 

\begin{equation*}
= \left\{ \begin{array}{ll}
            (1 - \beta_{r})^{(j)}P_1(r,i)\beta_{r} & \mbox{ $(i,j)\neq (N,0)$ }\\
         0 &\mbox{ $(i,j) = (N,0)$ }
			           		\end{array} \right. 
\end{equation*}\\
$P_{cP-T-R}^{(i,j)} $	& It is the probability that there is a collision between PS-POLL, TACK and RTS packets 
\begin{equation*}
= \left\{ \begin{array}{ll}
            P_1(r,j)P_1(r,i)\beta_{r} & \mbox{ $(i,j)\neq (N,0)$ }\\
         0 &\mbox{ $(i,j) = (N,0)$ }
			           		\end{array} \right. 
\end{equation*}\\

$P_{c}^{(i,j)}$ 			& It is the probability that there is a collision = $1 - (1 - \beta_{r})^{r} - r\beta_{r}(1 - \beta_{r})^{r-1} $

\label{tab:psm_prob_long_more_than_5}
\end{tabular}

\section{Mean fractions in different states} \label{appendix:fracs_times_PSM_long}

Let $E_{i,j}[S_{W_r}^l] = E_{i,j}\left[\int_{G_{k-1}}^{G_{k}} S_{W_r}(u)du\right] $.

\begin{align}
\begin{split}
&E_{i,j}[S_{W_r}^l] = P_{idle}^{(i,j)}[T_{W_k} + E_{i,j}[S_{W_r}^l]] + P_{sRTS}^{(i,j)}[T_{sRTS,W_k}] + P_{sTACK}^{(i,j)}T_{sTACK,W_k} + P_{sP}^{(i,j)}[T_{sP,W_k}] \\&+ \sum_{l=2}^{j}P_{cT,l}^{(i,j)}[T_{cT,W_k}^{(l)} + E_{i,j}[S_{W_r}^l]] + \sum_{l=1}^{j}P_{cT-R,l}^{(i,j)}[T_{cT-R,W_k}^{(l)} + E_{i,j}[S_{W_r}^l]]  \\& + \sum_{l=2}^{i}P_{cP,l}^{(i,j)}[T_{cP,W_K}^{(l)} + E_{i,j}[S_{W_r}^l]] + \sum_{l=1}^{i}P_{cP-R,l}^{(i,j)}[T_{cP-R,W_k}^{(l)} + E_{i,j}[S_{W_r}^l]] \\&+  \sum_{l=1}^{i}\sum_{m=1}^{j}P_{cT-P,(l,m)}^{(i,j)}[T_{cT-P,W_k}^{(l,m)} + E_{i,j}[S_{W_r}^l]] +  \sum_{l=1}^{i}\sum_{m=1}^{j}P_{cT-P-R,(l,m)}^{(i,j)}[T_{cT-P-R,W_k}^{(l,m)} + E_{i,j}[S_{W_r}^l]]\\
\end{split}
\end{align}
\begin{align}
\begin{split}
E_{i,j}[S_{W_r}^l] &= \frac{P_{idle}^{(i,j)}[T_{W_k}] + P_{sR}^{(i,j)}[T_{sR,W_k}] }{1 - P_{c}^{(i,j)} - P_{idle}^{(i,j)}} +\frac{P_{sT}^{(i,j)}T_{sT,W_k} + P_{sP}^{(i,j)}[T_{sP,W_k}] + \sum_{l=2}^{j}P_{cT,l}^{(i,j)}[T_{cT,W_k}^{(l)}]}{1 - P_{c}^{(i,j)} - P_{idle}^{(i,j)}}  \\&+\frac{  \sum_{l=1}^{j}P_{cT-R,l}^{(i,j)}[T_{cT-R,W_k}^{(l)} ] + \sum_{l=2}^{i}P_{cP,l}^{(i,j)}[T_{cP,W_K}^{(l)}]}{1 - P_{c}^{(i,j)} - P_{idle}^{(i,j)}}   \\&+\frac{ \sum_{l=1}^{i}P_{cP-R,l}^{(i,j)}[T_{cP-R,W_k}^{(l)} ] + \sum_{l=1}^{i}\sum_{m=1}^{j}P_{cT-P,(l,m)}^{(i,j)}[T_{cT-R,W_k}^{(l,m)}]}{1 - P_{c}^{(i,j)} - P_{idle}^{(i,j)}}   + \frac{\sum_{l=1}^{i}\sum_{m=1}^{j}P_{cT-P-R,(l,m)}^{(i,j)}[T_{cT-P-R,W_k}^{(l,m)}]}{1 - P_{c}^{(i,j)} - P_{idle}^{(i,j)}}
\end{split}
\end{align}
Notations used in the above equation are defined below and some are already being defined in Appendix~\ref{appendix:mean_cycle_length_long_psm_N>5} and they uses the 802.11 parameters defined in the Table~\ref{tab:param} 

\begin{tabular}{ p{2 cm}  p{12cm}}
$T_{Id,Id}$ & =$T_{Id,M_1}$ ,It is the total time spent by all the nodes in idle state when a system slot is idle $N\delta$\\
$T_{Id,Tx}$ &= $T_{Id,M_2}$ = 0\\
$T_{Id,Ls}$ &= $T_{Id,M_4}$ = 0\\
$T_{Id,RxD}$ &= $T_{Id,M_3}$ = 0\\
$T_{sR,Id}$  & It is the total time spent by all the nodes in idle when AP wins the contention = $N(3T_{SIFS} + T_{DIFS})$\\
$T_{sR,Tx}$  & It is the total time spent by all the nodes in Tx state when AP wins the contention = $T_{ACK} + T_{CTS}$\\
\end{tabular}

\begin{longtable}{ p{2 cm}  p{12cm}}
$T_{sR,Ls}$  & It is the total time spent by all the nodes in Ls state when AP wins the contention =  $(N-1)T_{DATA}$\\
$T_{sR,RxD}$ & It is the total time spent by all the nodes in RxD state when AP wins the contention =  $(N-1)[T_{ACK} + T_{CTS}] + NT_{RTS} + T_{DATA}$\\
$T_{sP,Id}$  & It is the total time spent by all the nodes in idle state when STA wins the contention and it transmits PS-POLL= $N(T_{SIFS} + T_{DIFS})$\\
$T_{sP,Tx}$  & It is the total time spent by all the nodes in Tx state when STA wins the contention and it transmits PS-POLL = $T_{PSPL}$ \\
$T_{sP,Ls}$  & It is the total time spent by all the nodes in Ls state when STA wins the contention and it transmits PS-POLL =$0$\\
$T_{sP,RxD}$  & It is the total time spent by all the nodes in RxD state when STA wins the contention and it transmits PS-POLL = $(N-1)T_{PSPL} + NT_{ACK}$\\
$T_{sT,Id}$  & It is the total time spent by all the nodes in idle state when STA wins the contention and it transmits TCP ACK = $N(T_{SIFS} + T_{DIFS})$\\
$T_{sT,Tx}$  & It is the total time spent by all the nodes in Tx state when STA wins the contention and it transmits TCP ACK = $T_{TACK}$\\
$T_{sT,LS}$  & It is the total time spent by all the nodes in Ls state when STA wins the contention and it transmits TCP ACK = $0$\\
$T_{sT,RxD}$  & It is the total time spent by all the nodes in RxD state when STA wins the contention and it transmits TCP ACK = $(N-1)T_{TACK} + NT_{ACK}$\\
$T_{cT,Id}^{(l)}$  & It is the total time spent by all the nodes in idle state when there is collision invloving $l$ STAs transmitting TCP ACK = $N T_{EIFS}$\\
$T_{cT,Tx}^{(l)}$  & It is the total time spent by all the nodes in Tx state when there is collision invloving $l$ STAs transmitting TCP ACK  = $lT_{TACK}$\\
$T_{cT,Ls}^{(l)}$  & It is the total time spent by all the nodes in Ls state when there is collision invloving $l$ STAs transmitting TCP ACK  = $0$\\
$T_{cT,RxD}^{(l)}$  &  It is the total time spent by all the nodes in RxD state when there is collision invloving $l$ STAs transmitting TCP ACK = $(N-l)T_{TACK} + NT_{ACK}$\\
$T_{cP,Id}^{(l)}$  & It is the total time spent by all the nodes in idle state when there is collision invloving $l$ STAs transmitting PS-POLL = $N T_{EIFS}$\\
$T_{cP,Tx}^{(l)}$  & It is the total time spent by all the nodes in Tx state when there is collision invloving $l$ STAs transmitting PS-POLL = $lT_{PSPL}$\\
$T_{cP,Ls}^{(l)}$  & It is the total time spent by all the nodes in Ls state when there is collision invloving $l$ STAs transmitting PS-POLL = $0$\\
$T_{cP,RxD}^{(l)}$  & It is the total time spent by all the nodes in RxD state when there is collision invloving $l$ STAs transmitting PS-POLL  =  $(N-l)T_{PSPL} + NT_{ACK}$\\
$T_{cT-R,Id}^{(l)}$ & It is the total time spent by all the nodes in idle state when there is collision invloving $l$ STAs transmitting TCP ACK and AP transmitting RTS = $N T_{EIFS}$\\
$T_{cT-R,Tx}^{(l)}$ & It is the total time spent by all the nodes in Tx state when there is collision invloving $l$ STAs transmitting TCP ACK and AP transmitting RTS = $l(T_{TACK})$ \\
$T_{cT-R,Ls}^{(l)}$ & It is the total time spent by all the nodes in Ls state when there is collision invloving $l$ STAs transmitting TCP ACK and AP transmitting RTS =   $0 $\\
$T_{cT-R,RxD}^{(l)}$ &It is the total time spent by all the nodes in RxD state when there is collision invloving $l$ STAs transmitting TCP ACK and AP transmitting RTS= $l(max(0, T_{RTS} - T_{TACK}) + (N-l)max(T_{RTS},T_{TACK})$\\
$T_{cP-R,Id}^{(l)}$ & It is the total time spent by all the nodes in idle state when there is collision invloving $l$ STAs transmitting PS-POLL and AP transmitting RTS= $N T_{EIFS}$\\
$T_{cP-R,Tx}^{(l)}$ & It is the total time spent by all the nodes in Tx state when there is collision invloving $l$ STAs transmitting PS-POLL and AP transmitting RTS= $l(T_{PSPL})$ \\
$T_{cP-R,Ls}^{(l)}$ & It is the total time spent by all the nodes in Ls state when there is collision invloving $l$ STAs transmitting PS-POLL and AP transmitting RTS=  $0 $ \\
$T_{cP-R,RxD}^{(l)}$ & It is the total time spent by all the nodes in RxD state when there is collision invloving $l$ STAs transmitting PS-POLL and AP transmitting RTS = $l(max(0, T_{RTS} - T_{PSPL}) + (N-l)max(T_{RTS},T_{PSPL})$\\
$T_{cP-T,Id}^{(l,m)}$ & It is the total time spent by all the nodes in idle state when there is collision invloving $l$ STAs transmitting PS-POLL and $m$ STAs transmitting TCP ACK = $N(T_{DIFS} + T_{EIFS})$\\
$T_{cP-T,Tx}^{(l,m)}$ &It is the total time spent by all the nodes in Tx state when there is collision invloving $l$ STAs transmitting PS-POLL and $m$ STAs transmitting TCP ACK = $lT_{PSPL} + mT_{TACK}$  \\
$T_{cP-T,Ls}^{(l,m)}$ & It is the total time spent by all the nodes in Ls state when there is collision invloving $l$ STAs transmitting PS-POLL and $m$ STAs transmitting TCP ACK  $0$ \\
%\end{tabular}
%\begin{longtable}{ p{2 cm}  p{12cm}}
$T_{cP-T,RxD}^{(l,m)}$ & It is the total time spent by all the nodes in RxD state when there is collision invloving $l$ STAs transmitting PS-POLL and $m$ STAs transmitting TCP ACK  = $l[max(0, T_{TACK} - T_{PSPL}] + m[max(0, T_{PSPL} - T_{TACK}] +  (N-l-m)[max(T_{PSPL},T_{TACK})]$\\
$T_{cP-T-R,Id}^{(l,m)}$ &It is the total time spent by all the nodes in idle state when there is collision invloving $l$ STAs transmitting PS-POLL, $m$ STAs transmitting TCP ACK and AP transmitting RTS = $N(T_{DIFS} + T_{EIFS})$\\
$T_{cP-T-R,Tx}^{(l,m)}$ & It is the total time spent by all the nodes in Tx state when there is collision invloving $l$ STAs transmitting PS-POLL, $m$ STAs transmitting TCP ACK and AP transmitting RTS = $lT_{PSPL} + mT_{TACK}$  \\
$T_{cP-T-R,Ls}^{(l,m)}$ & It is the total time spent by all the nodes in Ls state when there is collision invloving $l$ STAs transmitting PS-POLL, $m$ STAs transmitting TCP ACK and AP transmitting RTS = $0$  \\
$T_{cP-T-R,RxD}^{(l,m)}$ &It is the total time spent by all the nodes in RxD state when there is collision invloving $l$ STAs transmitting PS-POLL, $m$ STAs transmitting TCP ACK and AP transmitting RTS \,\,\,\,\,\,\,\,\,\,\,\,\,\,\,\,\,\,= $l[max(0, max(T_{TACK},T_{RTS}) - T_{PSPL}] + m[max(0, max(T_{PSPL},T_{RTS}) - T_{TACK}] +  (N-l-m)[max(T_{PSPL},T_{TACK}, T_{RTS})]$\\
\end{longtable}

\begin{tabular}{ p{2 cm}  p{12cm}}
$P_{cTACK,l}^{(i,j)}$ &  It is the probability that there is a collision between $l$ TCP ACK packets = ${^jC_l} \beta_r^{l}(1 - \beta_r)^{(r-l)}$\\
$P_{cPSPL,l}^{(i,j)}$ & It is the probability that there is a collision between $l$ PS-POLL packets  = ${^iC_l}\beta_r^{l}(1 - \beta_r)^{(r-l)}$\\
$P_{cP-R,l}^{(i,j)}$ &  It is the probability that there is a collision between $l$ PS-POLL packets and RTS =  ${^iC_l}\beta_r^{l+1}(1 - \beta_r)^{(i-l-1)} \,\,for\,\,(i,j) \neq (N,0)$\\
 &=	$0	\,\,for\,\,(i,j) = (N,0)$\\
$P_{cT-R,l}^{(i,j)}$ &  It is the probability that there is a collision between $l$ TCP ACK packets and RTS = ${^jC_l}\beta_r^{l+1}(1 - \beta_r)^{(j-l-1)}$ \\
$P_{cT-P-R,(l,m)}^{(i,j)}$ & It is the probability that there is a collision between $l$ PS-POLL, $m$ TCP ACK and RTS= ${^iC_l} {^jC_m}  \beta_r^{(l+m+1)} (1 - \beta_r)^{(r-l-m-1)}  \,\,for \,\,(i,j) \neq (N,0)$ \\
 &= $0 \,\, for\,\, (i,j) = (N,0)$ \\
\end{tabular}

\section{Mean cycle length - PSM LONG (N = 1)} \label{appendix:mean_cycle_length_long_psm_N1}
$E[T]$ is the expected time between two renewal instants, and is given by the following equation,

\begin{equation}
\begin{split}
 E[T] = &P_{id}(\delta + E[T]) + P_{sR}(T_{sR} + E[T_{PSPL}])  + P_{sT}T_{sT}  +  P_{cT-R}(T_{cT-R} + E[T]) \\
E[T] = &\frac{P_{id}\delta + P_{sR}T_{sR}  + P_{cT-R}T_{cT-R} }{1 - P_{idle} -  P_{c}}  +  \frac{P_{sT}T_{sT}}{1 - P_{idle} -  P_{c}}
\end{split}
 \end{equation}

\begin{equation}
\begin{split}
E[T_{PSPL}] & = \beta_2T_{sP} + (1 - \beta_2)(\delta + E[T_{PSPL}]) \\
E[T_{PSPL}] & = T_{sP} + \delta\frac{1 - \beta_2}{\beta_2}
\end{split}
\end{equation}
\\
Notations used in the above equation are defined previously in \ref{appendix:mean_cycle_length_long_psm_N>5} and below and uses they 802.11 parameters defined in the Table~\ref{tab:param}.
\\
\begin{tabular}{ p{2 cm}  p{12cm}}
$P_{id}$ & It is the probability of a slot being idle = $(1 - \beta_2)^2$\\
$P_{cT-R}$ & It is probability that there there is collision between TCP ACK and RTS= $\beta_2^2$\\
$P_{sR}$ & It is the probability of AP winning the contention = $\beta_2(1 - \beta_2)$\\
$P_{sT}$ & It is the probability of STA winning the contention =$\beta_2(1 - \beta_2)$ 
\end{tabular}

\section{Mean Cycle Length - Long tcp CAM } \label{appendix:mean_cycle_length_long_cam}
Expression for $E_k[T]$ can be written as follows,

\begin{align}
\begin{split}
&E_k[T] =   P_{idle,k}(\delta + E_k[T])  + \sum_{s=1}^{min(k,N)}P_{cR-T,k}^{(s)}(max(T_{RTS},T_{TACK}) + T_{EIFS} + E_k[T]) \\&+ 
\sum_{s=2}^{min(k,N)}P_{cT,k}^{(s)}(T_{TACK} + T_{EIFS} + E_k[T])  + P_{sT,k}(T_{SIFS} + T_{DIFS} + T_{TACK} + T_{ACK})  \\&+ P_{sR,k}(3T_{SIFS} + T_{DIFS} + T_{DATA} + T_{ACK})
\end{split}
\label{eqn:cam_long_mean_cycle_lt_1}
\end{align}

\begin{equation}
\begin{split}
&E_k[T] = \frac{P_{idle,k}\delta +  \sum_{s=1}^{min(k,N)}P_{cR-T,k}^{(s)}(max(T_{RTS},T_{TACK}) + T_{EIFS})}{1-P_{idle,k}- P_{c,k}}  \\&+ \frac{  \sum_{s=2}^{min(k,N)}P_{cT,k}^{(s)}(T_{TACK} + T_{EIFS}) + P_{sT,k}(T_{SIFS} + T_{DIFS} + T_{TACK} + T_{ACK})}{1-P_{idle,k}- P_{c,k}} +  \\&  + \frac{P_{sR,k}(3T_{SIFS} + T_{DIFS} + T_{DATA} + T_{ACK})}{1-P_{idle,k}- P_{c,k}}
\end{split}
\end{equation}

%\begin{tabular}{ l  {6.5cm}}
\begin{tabular}{ l  l}
$P_{idle,k}$& It is the probability of slot being idle \\
$P_{cR-T,k}^s$& It is the probability of collsion of $s$ TCP ACK and RTS\\
$P_{cT,k}^s$& It is the probability of collsion of $s$ TCP ACK only\\
$P_{c,k}$& It is the probability of collision\\
$P_{sR,k}$& It is the probability of success of AP\\
$P_{sT,k}$& It is the probability of success of STA\\
\end{tabular}

All the above probabilities can be expressed as following,

\begin{equation}
  P_{idle,k} = \left\{ \begin{array}{ll}
          (1-\beta_{k+1})^{k+1} & \mbox{for $0 \leq k < N$ };\\
         (1-\beta_{N+1})^{N+1} & \mbox{for $N \leq k \leq NW-1$ };\\
			(1-\beta_{N})^N  & \mbox{for $k = NW$ }.		\end{array} \right. 
\end{equation}

\begin{equation*}
  P_{cR-T,k}^{(s)} = \left\{ \begin{array}{ll}
			     0    & \mbox{for $k=0$}; \\
          \binom{k}{s}(\beta_{k+1})^{s+1}(1-\beta_{k+1})^{k-s} & 	\mbox{for $1 \leq k < N$ };\\
         \binom{N}{s}(\beta_{N+1})^{s+1}(1-\beta_{N+1})^{N-s} & \mbox{for  $N \leq k \leq NW-1$ };\\ 			0  & \mbox{for   $k = NW$ }.		\end{array} \right. 
\end{equation*}

\begin{equation*}
  P_{cT,k}^{(s)} = \left\{ \begin{array}{ll}
          0    & \mbox{for $0 \leq k \leq 1 $ }; \\
          \binom{k}{s}(\beta_{k+1})^{s}(1-\beta_{k+1})^{k-s+1} & \mbox{for $2 \leq k < N$ };\\
\binom{N}{s}(\beta_{N+1})^{s}(1-\beta_{N+1})^{N-s+1} & \mbox{for  $N \leq k \leq NW-1$ };\\
		\binom{N}{s}(\beta_{N})^{s}(1-\beta_{N})^{N-s}  & \mbox{for   $k = NW$ }.	\\
	\end{array} \right. 
\end{equation*}

\begin{equation}
P_{c,k} = \left\{ \begin{array}{lr}
          1-[(k+1)\beta_{k+1}(1-\beta_{k+1})^k) + (1-\beta_{k+1})^{k+1}] &  \mbox{for $0 \leq k < N$ }; \\
         1-[(N+1)\beta_{N+1}(1-\beta_{N+1})^N + (1-\beta_{N+1})^{N+1}]  & \mbox{for  $N \leq k \leq NW-1$ }; \\
			1-[(N)\beta_{N}(1-\beta_{N})^{N-1} + (1-\beta_{N})^{N}]  & \mbox{for   $k = NW$ }.		\end{array} \right. 
\end{equation}

\begin{equation}
P_{sR,k} = \left\{ \begin{array}{ll}
            \beta_{k+1}(1-\beta_{k+1})^{k} & \mbox{for $0 \leq k < N$ };\\
         \beta_{N+1}(1-\beta_{N+1})^{N} &\mbox{for  $N \leq k \leq NW-1$ };\\
			            0 & \mbox{for   $k = NW$ }.		\end{array} \right. 
\end{equation}

\begin{equation}
P_{sT,k} = \left\{ \begin{array}{ll}
            k\beta_{k+1}(1-\beta_{k+1})^{k} & \mbox{for $0 \leq k < N$ };\\
         N\beta_{N+1}(1-\beta_{N+1})^{N} &\mbox{for  $N \leq k \leq NW-1$ };\\
			N\beta_{N}(1-\beta_{N})^{N-1} & \mbox{for   $k = NW$ }.		\end{array} \right. 
\end{equation}

\section{Fraction of Times - Long tcp CAM } \label{appendix:fracs_times_CAM_long}
Let $E_{k}\left[\int_{G_{k-1}}^{G_{k}}S_{W_r}\right] = E_{k}[S_{W_r}^1]$. 

Expression for $E_{k}[S_{W_r}^1]$ can be written as follows,
\begin{equation*}
\begin{split}
&E_{k}[S_{W_r}^1] =  P_{idle,k}(E_{k}[S_{W_r}^1]  + T_{Id,W_r})+ \sum_{s=1}^{min(k,N)}P_{cR-T,k}^{(s)}(E_{k}[S_{W_r}^1] + T_{cAP,W_r}) \\&+ \sum_{s=2}^{min(k,N)}P_{cT,k}^{(s)}(E_{k}[S_{W_r}] + T_{cSTA,W_r}) + 
P_{sR,k}(T_{sAP,W_r} ) +  P_{sT,k}T_{sSTA,W_r}
\end{split}
\end{equation*}

\begin{equation}
\begin{split}
E_{k}[S_{W_r}^1] = &  \frac{P_{idle,k}T_{Id,W_r} + \sum_{s=1}^{min(k,N)}P_{cR,k}^{(s)}T_{cR-T,W_r} + \sum_{s=2}^{min(k,N)}P_{cT,k}^{(s)}(T_{cT,W_r}) + P_{sR,k}T_{sR,W_r}^1 }{1 - P_{idle,k} - P_{c,k} }  \\&+ 
 \frac{ P_{sT,k}T_{sT,W_r}}{1 - P_{idle,k} - P_{c,k}}
\end{split}
\end{equation}

Notations used in the above equation are defined below and Appenidix ~\ref{appendix:mean_cycle_length_long_cam}. They uses the 802.11 parameters defined in the Table~\ref{tab:param}

\begin{tabular}{ p{2 cm}  p{12cm}}
$T_{Id,Id}$ & =$T_{Id,M_1}$ ,It is the total time spent by all the nodes in idle state when a system slot is idle $N\delta$\\
$T_{Id,Tx}$ &= $T_{Id,M_2}$ = 0\\
$T_{Id,Ls}$ &= $T_{Id,M_4}$ = 0\\
$T_{Id,RxD}$ &= $T_{Id,M_3}$ = 0\\
$T_{sR,Id}$  & It is the total time spent by all the nodes in idle when AP wins the contention = $N(3T_{SIFS} + T_{DIFS})$\\
$T_{sR,Tx}$  & It is the total time spent by all the nodes in Tx state when AP wins the contention = $T_{ACK} + T_{CTS}$\\
$T_{sR,Ls}$  & It is the total time spent by all the nodes in Ls state when AP wins the contention =  $(N-1)T_{DATA}$\\
$T_{sR,RxD}$ & It is the total time spent by all the nodes in RxD state when AP wins the contention =  $(N-1)[T_{ACK} + T_{CTS}] + NT_{RTS} + T_{DATA}$\\
$T_{sT,Id}$  & It is the total time spent by all the nodes in idle state when STA wins the contention and it transmits TCP ACK = $N(T_{SIFS} + T_{DIFS})$\\
$T_{sT,Tx}$  & It is the total time spent by all the nodes in Tx state when STA wins the contention and it transmits TCP ACK = $T_{TACK}$\\
\end{tabular}

\begin{tabular}{ p{2 cm}  p{12cm}}
$T_{sT,LS}$  & It is the total time spent by all the nodes in Ls state when STA wins the contention and it transmits TCP ACK = $0$\\
$T_{sT,RxD}$  & It is the total time spent by all the nodes in RxD state when STA wins the contention and it transmits TCP ACK = $(N-1)T_{TACK} + NT_{ACK}$\\
$T_{cT,Id}^{(l)}$  & It is the total time spent by all the nodes in idle state when there is collision invloving $l$ STAs transmitting TCP ACK = $N T_{EIFS}$\\
$T_{cT,Tx}^{(l)}$  & It is the total time spent by all the nodes in Tx state when there is collision invloving $l$ STAs transmitting TCP ACK  = $lT_{TACK}$\\
$T_{cT,Ls}^{(l)}$  & It is the total time spent by all the nodes in Ls state when there is collision invloving $l$ STAs transmitting TCP ACK  = $0$\\
$T_{cT,RxD}^{(l)}$  &  It is the total time spent by all the nodes in RxD state when there is collision invloving $l$ STAs transmitting TCP ACK = $(N-l)T_{TACK} + NT_{ACK}$\\
\end{tabular}

\begin{tabular}{ p{2 cm}  p{12cm}}
$T_{cR-T,Id}^{(l)}$ & It is the total time spent by all the nodes in idle state when there is collision invloving $l$ STAs transmitting TCP ACK and AP transmitting RTS = $N T_{EIFS}$\\
$T_{cR-T,Tx}^{(l)}$ & It is the total time spent by all the nodes in Tx state when there is collision invloving $l$ STAs transmitting TCP ACK and AP transmitting RTS = $l(T_{TACK})$ \\
$T_{cR-T,Ls}^{(l)}$ & It is the total time spent by all the nodes in Ls state when there is collision invloving $l$ STAs transmitting TCP ACK and AP transmitting RTS =   $0 $\\
$T_{cR-T,RxD}^{(l)}$ &It is the total time spent by all the nodes in RxD state when there is collision invloving $l$ STAs transmitting TCP ACK and AP transmitting RTS= $l(max(0, T_{RTS} - T_{TACK}) + (N-l)max(T_{RTS},T_{TACK})$\\
\end{tabular}
\\
\\

\begin{table}
\renewcommand{\arraystretch}{1.3}
\caption{Transmission Times}
\begin{center}
\begin{tabular}{|l|l|p{5 cm}|}
\hline
\multicolumn{3}{|c|}{Transmission Times of Various Frames} \\ \hline
TCP-ACK  & $T_{TACK}$ & $T_P$ + $T_{PHY}$ + $\frac{L_{MAC}+L_{IPH}+L_{TACK}}{C_d}$ \\
MAC ACK &$T_{ACK}$   & $T_P$ + $T_{PHY}$ + $\frac{L_{ACK}}{C_c}$\\
PS-POLL &$T_{PSPL}$   & $T_P$ + $T_{PHY}$ + $\frac{L_{PSPL}}{C_c}$\\
TCP DATA &$T_{DATA}$   & $T_P$ + $T_{PHY}$ + $\frac{L_{MAC}+L_{IPH}+L_{TCPH}+L_{DATA}}{C_d}$\\ 
\hline
\end{tabular} \label{tab:param1}
\end{center}
\end{table}

\section{Average Current for 1 STA in PSM} \label{appendix:avg_current_psm_N1}
\subsection{$N = 1$ - Aggregate download throughput}~\label{section:appendix_long_file_psm_N1}
  After the Slow Start phase is over and the TCP window has grown to its maximum value, there will always be some packets at the AP and STA, with high probability. Due to this, the AP will always set the More bit in every outgoing packet; so STA will never go to sleep.
 
 When an STA receives a packet with the More bit set, it has to send a PS-POLL and a TCP ACK. PS-POLL, being MAC level packet, will be enqueued at the HOL position of the NIC queue, while the TCP ACK is enqueued at the end of the queue. 
 Since after the transmission of packet the AP queue is empty, so transmission of PS-POLL occurs without contention. However, TCP ACKs and data packets contend for transmission.
 
 Consider the process $X(t)$ denoting the number of TCP ACKs with the STA. The number of data packets with the AP is  $W - X(t)$. Denote the end of the $k^{th}$ success instants as $G_k$. Let $X_k$ be the number of TCP ACKs with the STA at $G_k$. Let $T_k = G_{k+1} - G_k$. Let the number of successful attempts by AP be $H(t)$ in time interval $(0,t)$. The number of successful attempts by the AP is 0 or 1 in between $G_k$ and $G_{k+1}$ with probability $0.5$. Then, using the Renewal Reward Theorem, the  following can be written,
 \begin{equation}
  \Theta_1 = \lim_{t \to \infty} \frac{H(t)}{t}= \frac{0.5}{E[T_k]}
 \end{equation}
 where, the detailed expression for $E[T]$ is given in Appendix~\ref{appendix:mean_cycle_length_long_psm_N1}. 

 \begin{table}
 \renewcommand{\arraystretch}{1.4}
 \caption{For N = 1}
 \label{tab:for_N_1}
 \centering
 \begin{tabular}{|c|c|c|c|c|c|c|}
 \hline
 &\multicolumn{3}{c|}{$\Theta_1$ (Mbps)} &\multicolumn{3}{c|}{$J_1$ (mA)} \\ \hline
             & 2     &   5.5  & 11      & 2        & 5.5     & 11 \\ \hline
   Analysis  & $1.28$ & $2.33$ & $3.04$  & $187.86$   & $197.37$  & $203.78$  \\\hline
 Simulation  & $1.22$ & $2.11$ & $2.66$  & $182.75$     & $185.15$ &  $ 186.45$                          \\ \hline
 \end{tabular}
 \label{tab:psm_single_long}
 \end{table}

Since in this case, there are only three possible states, which are $M_1$ = Idle State, $M_2$ = Tranmission State $M_4$ = Receive And Decode State.

\begin{equation}
J_{av,p} = J_{RxD} \Phi_{RxD} + J_{Tx} \Phi_{Tx} + J_{Id} \Phi_{Id}
\end{equation}

\begin{equation}
\begin{split}
\Phi_{W_r} = \frac{E[W_r]}{E[T_k]} \\
\end{split}
\end{equation}
\\

\begin{equation}
\begin{split}
&E[Id] =    P_{id}(\delta + E[Id]) + P_{cR-T}(T_{EIFS} + E[Id]) +  P_{sT,k}(T_{SIFS} + T_{DIFS}) + P_{sR,k}(4T_{SIFS} + 2T_{DIFS} + \delta\frac{1 - \beta_2}{\beta_2})\\
&E[Id]= \frac{P_{id}\delta + P_{cR-T}T_{EIFS}}{1 - P_{id} -  P_{c}} +  \frac{P_{sT}(T_{SIFS} + T_{DIFS}) + P_{sR}(4T_{SIFS} + 2T_{DIFS} + \delta\frac{1 - \beta_2}{\beta_2})}{1-P_{id}- P_{c}}
\end{split}
\end{equation}

\begin{equation}
\begin{split}
&E[RxD] \\&=  P_{id}E[RxD] + P_{cR-T}[max(0,T_{RTS}-T_{TACK}) + E[RxD]]   + P_{sT} T_{ACK} + P_{sR}(T_{RTS} + T_{DATA} + T_{ACK})\\
&E[RxD] \\&= \frac{P_{cR-T}max(0,T_{RTS}-T_{TACK})}{1-P_{id}- P_{c}} + \frac{P_{sT}T_{ACK} + P_{sR} (T_{RTS} + T_{DATA} + T_{ACK} )}{1-P_{id}- P_{c}}
\end{split}
\end{equation}

\begin{equation}
\begin{split}
E[Tx] = & P_{id}( 0 + E[Tx]) + P_{c}(T_{TACK} +  E[Tx]) + P_{sT} T_{TACK} +  P_{sR}(T_{CTS} + T_{ACK} + T_{PSPL})\\
E[Tx] = & \frac{P_{c}T_{TACK} + P_{sT} T_{TACK} }{1 - P_{id} - P_c} + \frac{ P_{sR}(T_{CTS} + T_{ACK} + T_{PSPL})}{1 - P_{id} - P_c}
\end{split}
\end{equation}

Notations used in the above equation are in Appenidix ~\ref{appendix:fracs_times_PSM_long} and \ref{appendix:mean_cycle_length_long_psm_N1}. They uses the 802.11 parameters defined in the Table~\ref{tab:param}. Results for the single STA in PSM downloading a large file over TCP is shown in Tab.~\ref{tab:psm_single_long}. There is a slight mismatch between the analysis and the simulation values. As discussed earlier STA goes to sleep state when there are no packets at the AP. When a single STA is downloading file then, then the partial TCP window remains at STA in the form of TCP ACKs and remaining at the AP in the form of data packets. If at any instant  whole TCP window comes in the form of TCP ACKs at the STA, it goes to sleep state, since the last packet it received must have More bit unset. STA remains in sleep state till the arrival of the next beacon frame, this results in lesser throughput and current values than the analytical values.

\section{Passive Current} \label{appendix:passive_current}
Since in this case STAs are only listening to the transmission not transmitting, so there are only three possible states, which are $M_1$ = Idle State, $M_3$ = Listen State, $M_4$ = Receive And Decode State.

\begin{equation}
J_{av,p} = J_{RxD} \Phi_{RxD} + J_{Ls} \Phi_{Ls} + J_{Id} \Phi_{Id}
\end{equation}

\begin{equation}
\begin{split}
\Phi_{W_r} &= \frac{\sum_{k=0}^{min(N,k)} \pi_k E_k[W_r]}{\sum_{k=0}^{min(N,k)} \pi_k E_k[X]} \\
\end{split}
\end{equation}
\\
\subsection{Calculation of expectations for passive node}

\begin{equation}
\begin{split}
&E_k[Id] =    P_{idle,k}(\delta + E_k[Id]) + P_{c,k}(T_{EIFS} + E_k[Id]) +   P_{sT,k}(T_{SIFS} + T_{DIFS}) + P_{sR,k}(3T_{SIFS} + T_{DIFS})\\
&E_k[Id]= \frac{P_{idle,k}\delta + P_{c,k}T_{EIFS}}{1 - P_{idle,k} -  P_{c,k}} +  \frac{P_{sT,k}(T_{SIFS} + T_{DIFS}) + P_{sR,k}(3T_{SIFS} + T_{DIFS})}{1-P_{idle,k}- P_{c,k}}
\end{split}
\end{equation}

\begin{equation}
\begin{split}
E_k[RxD] &=  P_{idle,k}E_k[RxD] + P_{cR-T,k}[max(T_{RTS},T_{TACK}) + E_k[RxD]] + 
P_{cT,k}(T_{TACK} + E_k[RxD])   \\&+ P_{sT,k}(T_{TACK} + T_{ACK}) + P_{sR,k}(T_{RTS} + T_{CTS} + T_{ACK})\\
E_k[RxD] &= \frac{P_{cR-T,k}max(T_{RTS},T_{TACK}) + P_{cT,k}T_{TACK}}{1-P_{idle,k}- P_{c,k}} \\&+ \frac{P_{sT,k}(T_{TACK} + T_{ACK}) }{1-P_{idle,k}- P_{c,k}}  + \frac{P_{sR,k}(T_{RTS} + T_{CTS} + T_{ACK} )}{1-P_{idle,k}- P_{c,k}}
\end{split}
\end{equation}

\begin{equation}
\begin{split}
E_k[Ls] = & P_{idle,k}( 0 + E_k[Ls]) + P_{c,k}(0 +  E_k[Ls]) + P_{sR,k}(T_{DATA})\\
E_k[Ls] =& \frac{P_{sR,k}T_{DATA}}{1-P_{idle,k}- P_{c,k}}
\end{split}
\end{equation}

Notations used in the above equation are in Appenidix ~\ref{appendix:mean_cycle_length_long_cam} and \ref{appendix:fracs_times_CAM_long}. They uses the 802.11 parameters defined in the Table~\ref{tab:param}


\begin{thebibliography}{1}
\bibitem{simulator:ns_2}
``The network simulator ns2.'',
  \url{http://www.isi.edu/nsnam/ns/.}
%
\bibitem{Analytical:processor_sharing_boucherie_2003}
R.~Litjens, F.~Roijers, J.~van~den Berg, R.~Boucherie, and M.~Fleuren,
  ``Performance analysis of wireless lans: an integrated packet/flow level
  approach,'' Enschede, 2003, imported from MEMORANDA.
  \url{http://doc.utwente.nl/65861/}

\bibitem{Analytical:processor_sharing_altman_2005}
D.~Miorandi, A.~A. Kherani, and E.~Altman, ``A queueing model for http traffic
  over ieee 802.11 wlans,'' \emph{Computer Networks}, vol.~50, no.~1, pp. 63 --
  79, 2006.
%
 \bibitem{Analytical:saving_energy_psm_pspoll_anastasi_04}
 G.~Anastasi, M.~Conti, E. Gregori and A.~Passarella, ``Saving energy in wi-fi
   hotspots through 802.11 psm: An analytical model,'' in \emph{2nd Workshop on
   Modeling and Optimization in Mobile, Ad Hoc and Wireless Networks
   (WiOpt'04)}, March, 2004, pp. 227--236.
% 
 \bibitem{Analytical:bulk_service_lei_nilsson_05}
 H.~Lei and A.~A. Nilsson, ``An m/g/1 queue with bulk service model for power
   management in wireless lans,'' in \emph{PE-WASUN '05: Proceedings of the 2nd
   ACM international workshop on Performance evaluation of wireless ad hoc,
   sensor, and ubiquitous networks}.\hskip 1em plus 0.5em minus 0.4em\relax New
   York, NY, USA: ACM, 2005, pp. 92--98.
% 
 \bibitem{Analytical:perform_extension_of_nelson_09}
 S.~Baek, B.~D.~Choi, ``Performance analysis of power save mode in ieee 802.11
   infrastructure wireless local area network,'' \emph{Journal of Industrial and
   Management Optimization (JIMO)}, vol.~5, no.~3, pp. 481--492, August, 2009.
%%
 \bibitem{Analytical:novel_down_link_scheme_pengbo_08}
 P.~Si, H.~Ji, F.~R.~Yu, G.~Yue, `Ieee 802.11 dcf psm model and a novel
   downlink access scheme,'' in \emph{Wireless Communications and Networking
   Conference, 2008. WCNC 2008. IEEE}, March 31 2008-April 3 2008, pp.
   1397--1401.
% 
 \bibitem{Analytical:micro_ibss_rong_06}
 R.~Zheng, J.~C, C.~Hou, and L.~Sha, ``A microscopic study of power
   management in ieee 802.11 wireless networks,'' \emph{Int. J. Wire. Mob.
   Comput.}, vol.~1, no. 3/4, pp. 165--178, 2006.
% 
 \bibitem{Experimental:ronny_bsd_05}
 R.~Krashinsky, H.~Balakrishnan, ``Minimizing energy for wireless web access
   with bounded slowdown,'' \emph{Wirel. Netw.}, vol.~11, no. 1-2, pp. 135--148,
   2005.
% 
 \bibitem{Experimental:smart_psm_quaio_05}
 D.~Qiao, Kang G.~Shin, ``Smart power-saving mode for ieee 802.11 wireless lans,'' in
   \emph{INFOCOM 2005. 24th Annual Joint Conference of the IEEE Computer and
   Communications Societies. Proceedings IEEE}, 2005, pp. 73-- 1583 vol. 3.
  \bibitem{Experimental:selftuning_night_03}
  M.~Anand, E.~B. Nightingale, J.~Flinn, ``Self-tuning wireless network power
   management,'' in \emph{MobiCom '03: Proceedings of the 9th annual
   international conference on Mobile computing and networking}.\hskip 1em plus
   0.5em minus 0.4em\relax New York, NY, USA: ACM, 2003, pp. 176--189.
% 
 \bibitem{Experimental:scheduled_psm_yong_07}
 H.~Yong, R.~Yuan, X.~Ma, J.~Li, C.~Wang, ``Scheduled psm for minimizing
   energy in wireless lans,'' in \emph{ICNP 2007.}, 16-19 Oct. 2007, pp.
   154--163.
% 
 \bibitem{Experimental:psm_throttling_enhua_07}
 E.~Tan, L.~Guo, S.~Chen, X.~Zhang, ``Psm-throttling: Minimizing energy consumption for
   bulk data communications in wlans,'' in \emph{ICNP 2007.}, 16-19 Oct. 2007,
   pp. 123--132.
 \bibitem{Analytical:math_analysis_dcf_04}
 A.~Zanella,  F.~D~.Pellegrini, ``Mathematical analysis of ieee 802.11 energy
   efficiency,'' in \emph{Proc. Int. Symp. The 7th International Symposium on
   Wireless Personal Multimedia Communications(WPMC2004)}, 12--15 September
   2004, pp. p.V1--97--p.V1--101.
% 
 \bibitem{Analytical:CAM_Nnodes_dharma_06}
 X.~Wang, J.~Yin, D.~P. Agrawal, ``Analysis and optimization of the energy
   efficiency in the 802.11 dcf,'' \emph{Mob. Netw. Appl.}, vol.~11, no.~2, pp.
   279--286, 2006.
% 
 \bibitem{Experimental:saving_energy_biamonte_06}
 V.~Baiamonte, C.-F. Chiasserini, ``Saving energy during channel contention
   in 802.11 wlans,'' \emph{Mob. Netw. Appl.}, vol.~11, no.~2, pp. 287--296,
   2006.
% 
 \bibitem{Base:kumar_fpa_05}
 A,~Kumar, E.~Altman, D.~Miorandi,   M.~Goyal,   ``New insights from a fixed-point analysis of
   single cell ieee 802.11 wlans,'' in \emph{IEEE INFOCOM '05}, March, 2005, pp.
   250--261.
% 
 \bibitem{Base:bcg_tcp_throughput_06}
 R.~Bruno, M.~Conti, and E.~Gregori, ``Performance modelling and measurements of
   tcp transfer throughput in 802.11-based wlan,'' in \emph{MSWiM '06:
   Proceedings of the 9th ACM international symposium on Modeling analysis and
   simulation of wireless and mobile systems}.\hskip 1em plus 0.5em minus
   0.4em\relax New York, NY, USA: ACM, 2006, pp. 4--11.
% 
 \bibitem{base:harsha_kuriakose}
 G.~Kuriakose, S.~Harsha, A.~Kumar, V.~Sharma, ``Analytical models for capacity
   estimation of ieee 802.11 wlans using dcf for internet applications,'' in
   \emph{Wireless Networks, Springer}, vol.~15, 2009, pp. 259--277 vol. 3.
%
\bibitem{book:book_rw_wolff}
R.~W. Wolff, \emph{Stochastic Modeling and the Theory of Queues
  (Paperback)}.\hskip 1em plus 0.5em minus 0.4em\relax Prentice Hall, 1989.
\bibitem{tech_spec:atheros_card_2state_power}
``Power consumption and energy efficiency comparisons of wlan products,''
  Atheros,
  \url{www.atheros.com/pt/whitepapers/atheros\_power\_whitepaper.pdf}
\bibitem{tech_spec:intel_card_current}
``Intel pro/wireless 2011 lan pc card,'' Intel.,
  \url{http://download.intel.com/support/wireless/wlan/pro2011/wireless.pdf}
%
\bibitem{rfc:http_persist}
``Hypertext transfer protocol -- http/1.1,'' 1999,
  \url{http://tools.ietf.org/html/rfc2616#section-5.1.1}
%
\bibitem{arxiv:our_paper_ana_mod}
``Analytical Models for Energy Consumption in Infrastructure WLAN STAs Carrying TCP Traffic'',
  \url{http://arxiv.org/abs/0909.3717v1} 
%
\bibitem{simulator:packmime}
``Packmime-http: Web traffic generation.'', 
  \url{http://www.isi.edu/nsnam/ns/doc/node552.html}
%
 \end{thebibliography}
\end{document}